\documentstyle[amsmath,amssymb,graphicx]{article}

\def\l[{\phantom.[}

\renewcommand{\arraystretch}{1.3}

\catcode`\@=11
\def\marginnote#1{}

\newcount\hour
\newcount\minute
\newtoks\amorpm
\hour=\time\divide\hour by60
\minute=\time{\multiply\hour by60 \global\advance\minute by-\hour}
\edef\standardtime{{\ifnum\hour<12 \global\amorpm={am}%
        \else\global\amorpm={pm}\advance\hour by-12 \fi
        \ifnum\hour=0 \hour=12 \fi
        \number\hour:\ifnum\minute<10 0\fi\number\minute\the\amorpm}}
\edef\militarytime{\number\hour:\ifnum\minute<10 0\fi\number\minute}

\def\draftlabel#1{{\@bsphack\if@filesw {\let\thepage\relax
      \xdef\@gtempa{\write\@auxout{\string
          \newlabel{#1}{{\@currentlabel}{\thepage}}}}}\@gtempa \if@nobreak
    \ifvmode\nobreak\fi\fi\fi\@esphack} \gdef\@eqnlabel{#1}}
    \def\@eqnlabel{}
\def\@vacuum{}
\def\draftmarginnote#1{\marginpar{\raggedright\scriptsize\tt#1}}

\def\draft{
  \oddsidemargin -.5truein
  \def\@oddfoot{\footnotesize \sl preliminary draft \hfil
    \rm\thepage\hfil\sl\today\quad\militarytime}
  \let\@evenfoot\@oddfoot \overfullrule 3pt
    \let\label=\draftlabel
    \let\marginnote=\draftmarginnote
  \def\@eqnnum{(\theequation)\rlap{\kern\marginparsep\tt\@eqnlabel}%
    \global\let\@eqnlabel\@vacuum}

  }

\makeatletter
\newdimen\normalarrayskip              
\newdimen\minarrayskip                 
\normalarrayskip\baselineskip
\minarrayskip\jot
\newif\ifold             \oldtrue            \def\new{\oldfalse}
\def\arraymode{\ifold\relax\else\displaystyle\fi} 
\def\eqnumphantom{\phantom{(\theequation)}}     
\def\@arrayskip{\ifold\baselineskip\z@\lineskip\z@
     \else
     \baselineskip\minarrayskip\lineskip2\minarrayskip\fi}
\def\@arrayclassz{\ifcase \@lastchclass \@acolampacol \or
\@ampacol \or \or \or \@addamp \or
   \@acolampacol \or \@firstampfalse \@acol \fi
\edef\@preamble{\@preamble
  \ifcase \@chnum
     \hfil$\relax\arraymode\@sharp$\hfil
     \or $\relax\arraymode\@sharp$\hfil
     \or \hfil$\relax\arraymode\@sharp$\fi}}
\def\@array[#1]#2{\setbox\@arstrutbox=\hbox{\vrule
     height\arraystretch \ht\strutbox
     depth\arraystretch \dp\strutbox
     width\z@}\@mkpream{#2}\edef\@preamble{\halign
\noexpand\@halignto
\bgroup \tabskip\z@ \@arstrut \@preamble \tabskip\z@ \cr}%
\let\@startpbox\@@startpbox \let\@endpbox\@@endpbox
  \if #1t\vtop \else \if#1b\vbox \else \vcenter \fi\fi
  \bgroup \let\par\relax
  \let\@sharp##\let\protect\relax
  \@arrayskip\@preamble}
\def\eqnarray{\stepcounter{equation}%
              \let\@currentlabel=\theequation
              \global\@eqnswtrue
              \global\@eqcnt\z@
              \tabskip\@centering
              \let\\=\@eqncr

 \halign to \displaywidth\bgroup
    \eqnumphantom\@eqnsel\hskip\@centering
    $\displaystyle \tabskip\z@ {##}$%
    \global\@eqcnt\@ne \hskip 2\arraycolsep
         $\displaystyle\arraymode{##}$\hfil
    \global\@eqcnt\tw@ \hskip 2\arraycolsep
         $\displaystyle\tabskip\z@{##}$\hfil
         \tabskip\@centering
    &{##}\tabskip\z@\cr}
\newfont{\hr}{msbm10}
\newfont{\ams}{msam10}

\hoffset= - 0.9in         

\def\beq{\begin{equation}}
\def\eeq{\end{equation}}
\def\ba{\beq\new\begin{array}{c}}
\def\ea{\end{array}\eeq}
\def\be{\ba}
\def\ee{\ea}

\def\N2{${\cal N}=2$}

\def\1N{${\cal N}=1$}
\def\4N{${\cal N}=4$}
\def\nn{\nonumber}

\newdimen\linethick  \linethick=0.4pt
\newdimen\hboxitspace    \hboxitspace=5pt
\newdimen\vboxitspace    \vboxitspace=5pt

\def\fr#1{%
\be
\vcenter{
\hrule height\linethick
          \hbox{\vrule width\linethick
                \kern\hboxitspace
                \vbox{\kern\vboxitspace
                      \hbox{$\begin{array}{c}\displaystyle#1
         \end{array}$}%
                      \kern\vboxitspace}%
                \kern\hboxitspace
                \vrule width\linethick}%
          \hrule height\linethick}%
\ee}

\hoffset= - 1.0in         

\begin{document}

\title{{\bf {On universal knot polynomials
}\vspace{.2cm}}
\author{{\bf A. Mironov$^{a,b,c,d,}$}\footnote{mironov@lpi.ru; mironov@itep.ru}, \ {\bf R. Mkrtchyan$^{e,}$}\thanks{mrl55@list.ru}, \ and \ {\bf A. Morozov$^{b,c,d,}$}\thanks{morozov@itep.ru},}
\date{ }
}

\maketitle

\vspace{-6.0cm}

\begin{center}
\hfill FIAN/TD-10/15\\
\hfill IITP/TH-13/15\\
\hfill ITEP/TH-24/15\\
\end{center}

\vspace{4.2cm}

\begin{center}
$^a$ {\small {\it Lebedev Physics Institute, Moscow 119991, Russia}}\\
$^b$ {\small {\it ITEP, Moscow 117218, Russia}}\\
$^c$ {\small {\it Institute for Information Transmission Problems, Moscow 127994, Russia}}\\
$^d$ {\small {\it National Research Nuclear University MEPhI, Moscow 115409, Russia }}\\
$^e$ {\small {\it Yerevan Physics Institute, 2 Alikhanian Br. Str., Yerevan 0036, Armenia}}
\end{center}

\vspace{1cm}

\begin{abstract}

We present a universal knot polynomials for 2- and 3-strand torus knots in adjoint representation, by universalization of appropriate Rosso-Jones formula. According to universality, these polynomials coincide with adjoined colored HOMFLY and Kauffman polynomials at SL and SO/Sp lines on Vogel's plane, and give their exceptional group's counterparts on exceptional line. We demonstrate that [m,n]=[n,m] topological invariance, when applicable, take place on the entire Vogel's plane. We also suggest the universal form of invariant of figure eight knot in adjoint representation, and suggest existence of such universalization for any knot in adjoint and its descendant representation. Properties of universal polynomials and applications of these results are discussed.
\end{abstract}

\vspace{.5cm}

\tableofcontents

\section{Introduction}

\subsection{Universality in simple Lie algebras and gauge theories}

 Representation theory is in the basis of our understanding of symmetries of
physical theories and plays an increasingly important role with revealing of
new hidden symmetries, sometime quite involved, which govern
the structure of states and dynamics of string theory and its field theory
reductions.

Representation theory looks different for different simple Lie algebras,
however, it seems that this is only because representations are classified by weights, not by roots.
The sub-sector of representation theory associated with roots
now shows significant signs of {\it universality}: many group-invariant quantities can be represented as values of analytical functions, defined over entire Vogel's plane (see the definition below), at a special points from Vogel's Table (\ref{VogelT}), (\ref{VogelExc}).

This sector is formed by the adjoint representation and representations appearing in decomposition of its tensor
powers. That is the reason to nickname it a $E_8$-sector of representation
theory, because for the maximal exceptional algebra $E_8$,
where the fundamental representation coincides with the adjoint one,
it provides a complete description.

This picture is not established yet.
There are arguments and conjectures pro and contra,
and the picture itself should be further clarified.
However, the perspective of unification of all
simple Lie algebras
(or at least some sectors of their representation theory)
is so attractive, that it certainly deserves a detailed study.
The present paper presents a new arguments
in favor of universality, by universalization of some types of knot polynomials.

The term "universality" refers to the notion of the
"Universal Lie algebra" introduced by Vogel in \cite{V2}, see also \cite{more},
which, roughly speaking, was intended to be a model for all simple Lie algebras.
That idea was based on his study of (his introduced) $\Lambda$-algebra of three-leg Jacobi diagrams
\cite{Vogel}, acting on different spaces of diagram and aimed finally to construction of finite Vassiliev's invariants of knots. These works present first impressive examples of universal quantities, such as e.g. dimensions of adjoint and its descendant representations,  and provide a motivation for subsequent developments. Latter particularly includes the whole series of universal dimension formulae in simple Lie algebras \cite{LM}, and proof of of universality of many quantities in Chern-Simons gauge theory \cite{MV12}.

An additional support to the universality comes from the geometric engineering \cite{geoing,Klemm}:
there gauge groups are secondary entities emerging from particular singularities
of Calabi-Yau spaces, and it is unnatural for algebras of different series to appear
in a different way, they should rather possess a common, i.e. universal description. Note that  in connected to this field Kostant's paper \cite{Kostant} the ADE part of Vogel's Table appears already at 1984.

There are a lot of open questions in this field, e.g. its relations to the Langlands theory \cite{Lang,KW}. Particularly important is problem of existence of a universal form of corresponding duality, which is a basis of our understanding of S-duality in string theories.

\subsection{Universal knot polynomials}

As shown in \cite{MV12,RM}, in the Chern-Simons gauge theory \cite{CS} on the $3d$ sphere
{\it universal} are such quantities as central charge, perturbative and non-perturbative partition functions,
etc., and particularly unknotted Wilson average in adjoint representation.
In the present paper we discuss the problem of universality of adjoint Wilson averages for some knotted curves.
It is well-known \cite{W1}, that they are connected with knot polynomials \cite{knotpols}.

Knot polynomials are usually defined with the help of concrete types of simple Lie algebras:
best known are the HOMFLY and Kauffman polynomials,
associated respectively with $SU(N)$ and $SO(N)$.
The definition of these polynomials already implies an analytical continuation from
positive integer to generic values of $N$, and the universality
is a far-going generalization, which can unify some of the HOMFLY and Kauffman polynomials
(those which are colored by representations from the set of adjoint descendants)
into a single quantity.
It is worth mentioning that finite Vassiliev's  invariants \cite{Vassinv}, which were among initial aims of Vogel's study \cite{Vogel}, arise in perturbative expansions of knot polynomials \cite{pertCS}.

Given these expectations, it is a natural task to lift entire knot polynomials to the universal level,
i.e. to define them as functions of the above mentioned  parameters  in such a way,
that (at least) the known  HOMFLY and Kauffman polynomials arise for their corresponding values,
and, moreover, at other special values one gets polynomials for $Sp(N)$ and for the exceptional groups.
The purpose of this paper is to demonstrate that  for the adjoint-colored polynomials this is indeed possible,
at least for some classes of knots.
Namely, we find universal knot polynomials for $2-$ and $3$-strand torus knots,
when Rosso-Jones formula \cite{RJ,RJmore} is available for any representation of any simple Lie algebra,
moreover, in this case the Rosso-Jones formula itself can be made universal.
We also do so for the figure eight knot $4_1$, where this provides a new set of colored HOMFLY polynomials
and continuation to exceptional groups is a new result of its own value.
Remarkably, the universal formulas inherit distinguished properties
of ordinary knot polynomials like evolution \cite{DMMSS,evo}, factorization of special
polynomials \cite{DMMSS,IMMMfe,spepo} and differential expansion \cite{diffexpan}.

\bigskip

Actual development proceeded as follows.

First, for any given link/knot ${\cal L}$ we introduced  the "uniform adjoint HOMFLY"
${\cal H}_{\rm Adj}^{\cal L}(q|A)$, which are still polynomials in $A=q^N$ despite the appearance of N in parametrization of its Young diagram:
\be
{\cal H}^{\cal L}_{\rm Adj}(q|A=q^N)=H^{\cal L}_{[21^{N-2}]}(q|A=q^N)
\label{uniH}
\ee
This new polynomial in all calculated cases is remarkably simple and possesses wonderful properties.

Second, we consider the adjoint Kauffman polynomial
\be
{\cal K}^{\cal L}_{\rm Adj}(q|A) = K^{\cal L}_{[11]}(q|A)
\label{uniK}
\ee
(note that $A=q^{N-1}$ for $SO(N)$ and $Sp(-N)$). It is known for a limited (classes of) knots, however wide enough to establish universality in cases we interested in.

Third, for a few small knots we found a universal polynomial ${\cal U}^{\cal L}_{\rm Adj}(u,v,w)$,
which is a {\it symmetric} Laurent polynomial of
three variables $u=q^\alpha$, $v=q^\beta$ and $w=q^\gamma$ ($\alpha$, $\beta$ and $\gamma$
are the conventional Vogel's parameters \cite{Vogel}),
and interpolates between the uniform HOMFLY and Kauffman polynomials, when specialized to SU and SO lines of Table (\ref{VogelT}):\footnote{
For a different relation between colored Kauffman and HOMFLY polynomials see \cite{MarK}.
}

\be
{\cal U}^{\cal L}_{\rm Adj}(u=q^{-2},v=q^2,w=A) = {\cal H}^{\cal L}_{\rm Adj}(q|A),   \\
{\cal U}^{\cal L}_{\rm Adj}(u=q^{-2},v=q^4,w=q^{-3}A) = {\cal K}^{\cal L}_{\rm Adj}(q|A)
\label{interpo}
\ee
It is a highly non-trivial fact that such an interpolation exists at all: the corresponding system of equations on coefficients of the mentioned Laurent polynomial is highly overdetermined, but appears to have a solution. To understand a non-triviality of this fact, one can try to represent in such a way the simplest
fundamental HOMFLY polynomial
$H^{4_1}_{[1]} = 1+A^2+A^{-2}-q^2-q^{-2} = 1+\{Aq\}\{A/q\} $
nothing to say, to interpolate it to the
fundamental Kauffman polynomial
$
K^{4_1}_{[1]} = 1+(q^2-1+q^{-2})\cdot\Big((A^2+A^{-2})-(q-q^{-1})(A-A^{-1})-2\Big)
= 1 + (q^2-1+q^{-2})\cdot\frac{(Aq+1)(A-q)\{A\}}{Aq}
$.
One of the sources of problem is that any Kauffman polynomial is an invariant of non-oriented links only, while the fundamental HOMFLY polynomial is an invariant of
oriented links. However, the HOMFLY polynomial in the adjoint representation also does not differ between differently oriented links, since the adjoint representation is self-conjugated.

On the other hand, (\ref{interpo}) defines the universal polynomial (if it exists)
ambiguously: only modulo a symmetric polynomial
\be
(uv-1)(uw-1)(vw-1)(u^2v-1)(uv^2-1)(u^2w-1)(uw^2-1)(v^2w-1)(vw^2-1)
\label{ambig}
\ee
of a relatively high degree $24$.
For very small knots one can fix this ambiguity by requirement of having polynomial of minimal degree, but for
bigger knots with knot polynomials of higher degree this doesn't work.

Forth, we lifted formulas for small knots to those for the entire 1-parametric $2-$ and $3$-strand
torus families.
This lifting helps us to tame/reduce the
ambiguity for at least the knots of this type.

Fifth, finally we found a way to directly derive universal expressions for the 2- and 3-strand knots from the general Rosso-Jones formula, valid for any group. As original Rosso-Jones expression, it is valid (and hence prove universality) for corresponding links, also.

Below we concentrate mainly on the universal Rosso-Jones formula,
omitting most details of our original calculation.

However, original methods, and universality statement/hypothesis are actually applicable/suggested to an arbitrary knots, if sufficiently much is known about their colored HOMFLY
and Kauffman polynomials.
As an illustration, we present the answer for the universal adjoint polynomial
of the figure-eight knot $4_1$.
We stress again  that {\it this} answer should be used with a certain care,
because the ambiguity issue is not fully resolved.
The real resolution would come from lifting to the Vogel universal level of the full
Rosso-Jones formula and, more generally, of the modern version of the
Reshetikhin-Turaev formalism
\cite{RT}-\cite{knotebook}.

Finally, some properties of universal polynomials are discussed in the
last section.

\section{Adjoint polynomials \label{adjo}}

Adjoint is a distinguished representation from many points of view,
both in physics and in mathematics.
Closer to our purposes in this paper,
universal Vogel's \cite{Vogel} and Landsberg and Manivel's \cite{LM} formulas (see also \cite{more}) deal with (some) irreducible representations in an arbitrary powers of adjoint representation of Lie algebra,
while its extension to all representations, including fundamental, is problematic.
In this section we collect just some  of the relevant technical details.

\subsection{Rosso-Jones formula}

The original expression for invariants of torus knots/links, derived by Rosso and Jones in \cite{RJ},
looks as

\be
P^{[m,n]}_R = \frac{ q^{mn\varkappa_R}}{D_R(q)} \sum_Y \sum_Q
q^{-\frac{n}{m}\varkappa_{_Q}}  \varphi_{_Y} (\bar{\sigma}^{[m,n]}) D_{_Q}(q)
\label{RJ2}
\ee

Here $Y$ runs over all Young diagrams with m boxes, $Q$ runs over all irreducible representations, with multiplicities, of the gauge group (i.e. the group, of which $R$ is the representation) in one of the subspaces of $R^{\otimes m}$ with symmetry $Y$, $\varphi_{_Y} (\bar{\sigma}^{[m,n]})$ is the character of $Y$  representation of symmetric group $S_m$ evaluated on the element $(\sigma_1...\sigma_{m-1})^n$ of braid group $B_m$, considered as an element of $S_m$, $\varkappa_Q$ is the second Casimir of $Q$, $D_{_Q}(q)$ are the usual quantum dimensions, see below.  Elements $\sigma_1, ..., \sigma_{m-1}$ are usual generators of the braid group $B_m$, interchanging two neighboring strands, so their product reduced to the symmetric group $S_m$ is just a cyclic permutation $(1,2,...m)$. Obviously, its $m$-th power will be an identity element. If $m, n$ are mutually prime, this is a knot invariant, otherwise it is the invariant of the corresponding $l$-component link, where $l$ is the greatest common divisor of $m$ and $n$. The Rosso-Jones formula is given here in the topological framing.

The Rosso-Jones formula can be rewritten in a more inspiring form \cite{RJmore}

\be
P^{[m,n]}_R(q)\ = \ \frac{  q^{mn\varkappa_R}}{D_{\!_R}(q )} \sum_{Q\in R^{\otimes m}}
c_{_{RQ}} \cdot q^{-\frac{n}{m}\varkappa_{_Q}}\cdot D_{_Q}(q )
\label{RJ}
\ee
It also has far going generalizations to arbitrary knot polynomials
in braid realizations \cite{RTmod,RTnew}.

This formula treats differently the number $m$ of strands in the braid and its length
(evolution parameter) $n$: the $m\longleftrightarrow n$ symmetry of the answer,
$P^{[m,n]}_R=P^{[n,m]}_R$,  necessary for its topological invariance,
is technically a non-trivial fact.
The sum in this formula goes over all irreducible representations $Q$, belonging to the
$m$-th power of the original representation $R$,
\be
\varkappa_Q = (\Lambda_Q,\Lambda_Q+2\rho)
\ee
is the corresponding eigenvalue of the Casimir operator
and
\be\label{Weyl}
D_Q=\prod_{\alpha\in\Delta_+}{[\left(\Lambda_Q+\rho,\alpha\right)]\over [\left(\rho,\alpha\right)]}
\ee
is its quantum dimension.
Here $\Lambda_Q$ is the highest weight of the representation $Q$,
$\rho$ is the
Weyl vector, equal to the half sum of positive roots, and
square bracket denotes the quantum number:
\be
[x]=\frac{q^x-q^{-x}}{q-q^{-1}} = \frac{\{q^x\}}{\{q\}},\ \ \ \ \ \ \{x\}=x-x^{-1}
\ee
The coefficients $c_{RQ}$ are integers.

They are explicitly given by a somewhat sophisticated formula \cite{RJ}:
\be
c_{_{RQ}} =  \varphi_{_{W_Q}} (\bar{\sigma}^{[m,n]})
\label{cWQ}
\ee
i.e. $c_{RQ}$ is the corresponding value of symmetric group character $\varphi$,
in representation $W_Q$ of $S_m$, which describes the multiplicity of representation $Q$ in decomposition
\be
R^{\otimes m} = \oplus_Q \  W_Q\otimes Q
\ee

A more elegant version is to define $c_{_{RQ}}$ through characters of the original algebra,
extended to the space of time-variables (this is a well known procedure for $SU(N)$,
but requires a more detailed explanation in the case of exceptional algebras).
Then one can apply the Adams plethysm rule: for knots
\be
\hat Ad_m \chi_{_R}(p_k) \equiv \chi_{_R}(p_{mk}) = \sum_{Q\in R^{\otimes m}}
c_{_{RQ}} \chi_{_Q}(p_k)
\label{Adrule}
\ee
and
\be
\prod_{i=1}^l \hat Ad_{_{m/l}} \chi_{_{R_i}}  = \sum_Q c_{_{\vec RQ}} \chi_{_Q}(p_k)
\ee
for the $l$ component link in representation $\vec R = \otimes_{i=1}^l R_i$.
The quantum dimensions are restrictions of  time-dependent characters to the "topological locus":
\be
D_Q = {\rm tr}_{_Q} q^{\rho} =   \chi_Q(p_k^*),
\ \ \ \ \ \ \ \ p_k^* = {\rm tr}_{_\Box} q^{k\rho}
\ee
where the last trace is taken in the fundamental representation.
Clearly, these definitions imply an extension of the r.h.s. of (\ref{RJ}) to the entire space
of time variables \cite{MMMkn1}, however such $H^{[m,n]}_R(q|p_k)$ does not possess
the $m\longleftrightarrow n$ symmetry, i.e. is not fully topologically invariant
(depends also on the braid representation).
If one prefers to work entirely on the topological locus,
one should use the original (\ref{cWQ}).







\bigskip

For $SU(N)$ the knot polynomial $P$
is called HOMFLY-PT polynomial, and we denote it by $H$.
Usually HOMFLY is defined as a polynomial of two variables $q$ and $A$,
specialization to particular $SU(N)$ is provided by putting $A=q^N$.

For $SO(N)$ it is called Kauffman polynomial, denoted by $K$ and specialization is $A=q^{N-1}$.
The $Sp(N)$ case can be  obtained from the $SO(N)$ one by the substitution $N \longrightarrow -N$,  transposition of Young diagrams and renormalization of scalar product in algebra (or, equivalently, on the language of Chern-Simons theory, by renormalization of coupling constant), see \cite{SOSp,MV12} for gauge theories' side of this equivalence, and \cite{DMMSS,IMMMfe} for that in knots theory.
Some extra modifications are
needed in the case of superpolynomials, see \cite{DMMSS} and \cite{superpRJ}.

Isomorphisms between different small groups imply relations between HOMFLY and Kauffman polynomials
(for the purposes of this paper
we restrict formulas to adjoint representation, $Adj_{_{SO}}=[11]$, $Adj_{_{Sp}} = [11]^{tr}=[2]$). With appropriate choice of normalization of scalar products one has:
\be
\begin{array}{ccc}
SO(3)\cong SU(2)/Z_2 & \Longrightarrow &
{\cal K}^{ }_{Adj}(q^2|A=q^4) = {\cal H}^{ }_{Adj}(q|A=q^2)   \\
SO(4)\cong SU(2)^2/Z_2 & \Longrightarrow &
{\cal K}^{ }_{Adj}(q|A=q^3) = {\cal H}^{ }_{Adj}(q|A=q^2)   \\
SO(6)\cong SU(4)/Z_2 & \Longrightarrow &
{\cal K}^{ }_{Adj}(q|A=q^5) = {\cal H}^{ }_{Adj}(q|A=q^4)   \\
SU(2) \cong  Sp(2)  & \Longrightarrow &
{\cal H}^{ }_{Adj}(q^{2}|A=q^{4}) =
{\cal K}^{ }_{Adj}(q^{-1}|A=q^{3})     \\
SO(5) \cong Sp(4)/Z_2 & \Longrightarrow &
{\cal K}^{ }_{Adj}(q^{2}|A=q^{8}) =
{\cal K}^{ }_{Adj}(q^{-1}|A=q^{5})
\end{array}
\label{Sp4S05}
\ee
As a corollary,
\be
{\cal K}^{ }_{Adj}(q^{2}|A=q^{6}) = {\cal K}^{ }_{Adj}(q^{4}|A=q^{8}) =
{\cal K}^{ }_{Adj}(q^{-1}|A=q^{3})
\ee



Twist knots can be described by a very similar evolution formula \cite{evo,MMM21},
only in this case $m=2$, but $Q\in R\otimes \bar R$, where $\bar R$ is a conjugate representation,
and $c_{_{RQ}}$ are substituted by more complicated expressions, requiring separate tedious calculations.

\subsection{$SU(N)$ series}

For $SU(N)$ the parameter $A=q^N$ captures all the dependence on $N$, provided the quantum dimensions $D_{_Q}(q|A)$
are also expressed through it, see below. For arbitrary representation $R$ in this case, the second Casimir is equal to
\be\label{Cas}
\varkappa_R=2\kappa_R-{|R|^2\over N}+|R|N
\ee
with $\kappa_R=\sum_{r_{i,j}\in R}(j-i)$, where the sum goes over the boxes of the Young diagram $R$ and $\kappa_R$ is
the corresponding eigenvalue of the cut-and-join operator \cite{MMN1},
\be
\hat W_2 \chi_Q = \kappa_Q\chi_Q
\ee
Note that the shift of $\kappa_R$ in (\ref{Cas}) is essential in order to guarantee that $\varkappa_{[1^N]}= 0$, since representation $[1^N]$ is equivalent to the singlet.

For $SU(N)$, the adjoint representation is associated with the $N$-dependent self-conjugate hook diagram
\be
Adj = [21^{N-2}] \ \ \ \text{for} \ SU(N)
\ee
As usual, conjugate is the diagram, which after rotation by $\pi$ can be "added" to
the original diagram to form a full rectangular of the length/height $N$.

What also distinguishes adjoint representation of $SU(N)$ is the slow growth of its dimension
$d_{adj} = [N+1][N-1]$ for large $N$, which signals about strong cancelations in the hook formula
\be
D_R = \prod_{(i,j)\in R} \frac{[N+i-j]}{[{\rm hooklength}_{(i,j)}]}
\ee
(normally dimension of an $M$-box diagram would grow as $N^M$, but since the denominator can also
grow as fast as $M!$, compensation is possible for $M \sim N$, and it indeed takes place for the adjoint representation).
This also implies that the powers of adjoint representation decompose into a relatively small
number of irreducibles, just seven:
\be\label{adj2}
Adj^{\otimes 2} = [21^{N-2}]\otimes [21^{N-2}] = [42^{N-2}]\oplus \Big([42^{N-3}1^2]=[31^{N-3}]\Big)
\oplus [332^{N-3}]\oplus   \\
\oplus \Big([332^{N-4}1^2]=[221^{N-4}]\Big)
\oplus (2-\delta_{N,2})\cdot\Big([32^{N-2}1]=[21^{N-2}]\Big)
\oplus \Big([2^N]=[0]\Big)
\ee
where [0] denotes one-dimensional singlet representation.

Technically this works as follows: say, for $SU(4)$ adjoint representation is $[211]$,
and from the decomposition
\be
[211]^{\otimes 2} = \underline{[422]} \oplus
\Big(\underline{\underline{[4211]\stackrel{N=4}{\longrightarrow} [31]}}\Big)
\oplus [41111]
\oplus \underline{[332]}  \oplus
\Big(\underline{\underline{[3311]\stackrel{N=4}{\longrightarrow} [22]}}\Big) \oplus   \\
\oplus
2\cdot\Big(\underline{\underline{[3221]\stackrel{N=4}{\longrightarrow} [211]}}\Big) \oplus 2\cdot [32111]
\oplus[311111] \oplus
\Big(\underline{\underline{[2222]\stackrel{N=4}{\longrightarrow} [0]}}\Big) +[22211] + [221111]
\ee
only the six underlined Young diagrams (one with multiplicity two)
have no more than $N=4$ lines and survive for $SU(4)$,
moreover, the double-underlined diagrams with exactly $N=4$ lines are further simplified.

Note also that the adjoint representation is self-conjugate, and so are the five representations in
its square, the remaining two are conjugate of each other,  $\overline{[31^{N-3}]}=[332^{N-3}]$.
However, if we extend SU(N) group by automorphysms of its Dynkin diagram, then the sum of two last representations becomes one irreducible representation of extended group.
It also deserves mentioning that quantum dimensions of symmetric and antisymmetric squares of
the representation $R$ are equal to
\be
\frac{D_R^2 \pm \hat Ad_2 (D_R)}{2} \equiv \frac{D_R(q|A)^2 \pm D_R(q^2|A^2)}{2}
\ee
In particular, for $SU(N)$, i.e. at $A=q^N$ one has,
say,
\be
\frac{D_2(q)^2+D_2(q^2)}{2} = D_4(q|A)+D_{22}(q|A)
\ee

and

\be
\frac{D_{Adj}(q)^2 + D_{Adj}(q^2)}{2} = D_{[42^{N-2}]}(q)+D_{[221^{N-4}]}(q) + D_{Adj}(q)+1  \\
\frac{D_{Adj}(q)^2 - D_{Adj}(q^2)}{2} = D_{[332^{N-3}]}(q)+D_{[31^{N-3}]}(q)+D_{Adj}(q)
\ee
where the dimensions are:
\be
D_{[42^{N-2}]}(q)={\{A\}^2\{Aq^3\}\{A/q\}\over \{q^2\}^2\{q\}^2},\ \ \ \ \ \ \ \ \ D_{[221^{N-4}]}(q)={\{Aq\}\{A\}^2\{A/q^3\}\over
\{q^2\}^2\{q\}^2}, \\
D_{Adj}={\{Aq\}\{A/q\}\over \{q\}^2},\ \ \ \ \ \ \ \ \
D_{[332^{N-3}]}(q)= D_{[31^{N-3}]}(q)={\{Aq^2\}\{Aq\}\{A/q^2\}\{A/q\}\over\{q^2\}^2\{q\}^2}
\ee

\bigskip

The 2-Adams plethysm decomposition is even smaller than the square of the adjoint (\ref{adj2}): it does not contain the item with multiplicity two
(one symmetric and one anti-symmetric: they hence drop out of the alternated sum):
\be
\hat{Ad}_2(Adj) = [21^{N-2}]\otimes [21^{N-2}] = [42^{N-2}]\ominus \Big([42^{N-3}1^2]=[31^{N-3}]\Big)
\ominus [332^{N-3}]\oplus   \\
\oplus \Big([332^{N-4}1^2]=[221^{N-4}]\Big)
\ominus \delta_{N,2}\Big([31]=[2]\Big) \oplus \Big([2^N]=[0]\Big)
\ee
so that the Rosso-Jones formula for the adjoint-colored
HOMFLY polynomials of the 2-strand torus knots states:
\be\label{30}
{\cal H}^{[2,n]}_{Adj}(q|A) ={1\over D_{Adj}}\left[\left({A\over q}\right)^{2n}D_{[42^{N-2}]}+\left({A q}\right)^{2n}D_{[221^{N-4}]}-
A^{2n}\Big(D_{[31^{N-3}]}+D_{[332^{N-3}]}\Big)+A^{4n}\right]
\ee

\subsection{$SO/Sp$ series}

Similarly to the $SU(N)$ case, representations of $SO(N)$ are also labeled by Young diagrams, besides spinor ones.
The quantum dimensions $d_R$ for various representations of $SO(N)$ are very similar
to $D_R$ of $SU(N)$, if expressed through the parameter $A$, but with a notable change:
one has to parameterize $A=q^{N-1}$ in $d_R$ for $SO(N)$, instead of $A=q^N$ in $D_R$ for $SU(N)$. These dimensions can be calculated using formula (\ref{Weyl}) or \cite[(4.9)]{BFM} for the representation given by the Young diagram $R$ with the lines $\{r_i\}$,
$i=1,\ldots, l(R)$:
\be\label{qdSO}
d_R=\prod_{1\le i<j\le l(R)}{\{q^{r_i-r_j+j-i}\}\{Aq^{r_i+r_j+1-i-j}\}\over\{q^{j-i}\}\{Aq^{1-i-j}\}}\times
\prod_{k=1}^{l(R)}\left({\{(A^{1/2}q^{r_k-k+{1\over 2}}\}\over \{A^{1/2}q^{-k+{1\over 2}}\}}\prod_{s=1}^{r_k}{\{Aq^{r_k+1-k-s-l(R)}\}\over \{q^{s-k+l(R)}\}}
\right)
\ee
A table of the first dimensions and their product rules can be found in the Appendix A.

The Adams plethysm relations look like
\be
\hat Ad_{2}\Big(d_{[1]}\Big) = d_{[2]} - d_{[11]}+1,   \\
\hat Ad_3\Big( d_{[1]}\Big) = d_{[3]}-d_{[21]}+d_{[111]},   \\
\hat Ad_2\Big(d_{[2]}\Big) = d_{[4]}-d_{[31]}+d_{[22]} + d_{[2]}-d_{[11]}+1   \\
\hat Ad_2\Big(d_{[11]}\Big) = d_{[22]}-d_{[211]}+d_{[1111]} + d_{[2]}-d_{[11]}+1   \\
\ldots
\ee

The adjoint representation of $SO(N)$ is independent of $N$ and is just the first
antisymmetric representation:
\be
Adj=[11]  \ \ \ \ \text{for}\ SO(N)
\ee
Its symmetric and antisymmetric squares are decomposed as follows:
\be
{\cal S}^2 (Adj) = {\cal S}^2([11]) = [22]+[1111]+[2]+[0], \\
\Lambda^2 (Adj) = \Lambda^2([11]) = [211]+[11] = [211]+Adj
\ee
and, accordingly,
\be
\frac{d_{[11]}^2(q) + d_{11}(q^2)}{2} = d_{22}(q) + d_{1111}(q)+d_{2}(q)+1,   \\
\frac{d_{[11]}^2(q) - d_{11}(q^2)}{2} = d_{[211]}(q)+d_{[11]}(q)
\ee
Also note that unlike the $SU(N)$ case there are now only two irreducible
representations in the antisymmetric square, but one of them is still adjoint.
However, there is no longer an adjoint in the symmetric square.

Since the Young diagram of the adjoint representation for $SO(N)$ is the same for all $N$,
calculation of the adjoint Kauffman polynomial is much simpler.
According to \cite{Ste}, for the two-strand torus knots $[2,n]$ one has:
\be
{\cal K}^{[2,n]}_{Adj}= K^{[2,n]}_{[11]}=
\frac{q^{-4n}A^{4n}}{d_{[11]}}\left( A^{-2n}d_{[22]} - q^{2n}A^{-2n}d_{[211]}+q^{6n}A^{-2n}d_{[1111]}+
q^{-n}A^{-n}d_{[2]}-q^nA^{-n}d_{[11]}+1
\right)
\label{adjkau}
\ee
The fundamental and symmetric reduced Kauffman polynomials are respectively
\be
K^{[2,n]}_{[1]}=\frac{ A^{2n}\Big(q^{-n}A^{-n}d_{[2]} - q^n A^{-n}d_{[11]}+1\Big)}{d_{[1]}}
\ee
and
\be
K^{[2,n]}_{[2]}=\frac{q^{4n}A^{4n}}{d_{[2]}}
\left(q^{-6n}A^{-2n}d_{[4]} - q^{-2n}A^{-2n}d_{[31]}+A^{-2n}d_{[22]}+
q^{-n}A^{-n}d_{[2]}-q^nA^{-n}d_{[11]}+1
\right)
\ee
For comparison with the fundamental Kauffman polynomials in \cite{katlas}
one should substitute there $A\longrightarrow -iA$ and $z=i\{q\}$.


\subsection{Exceptional algebras}

 It is possible to calculate, in a similar fashion, quantum dimensions (see Appendix B) and knots polynomials for some knots (e.g. two-strand knots) for exceptional algebras. However, this is more time-consuming, and, more important, as we shall see in the next section, actually exceptional algebras have some similarity with classical series in a sense that they all are located on the line in Vogel's plane. So we present below polynomial for trefoil on that line, i.e. simultaneously for all exceptional algebras, as is done above for $SU$ and $SO/Sp$ lines.  Answer for particular algebras appear at special values of parameter $N$, given (as well as definitions of Vogel's plane, exceptional line, etc) in next Section, (\ref{VogelT}), (\ref{VogelExc}).

Adjoint knot polynomial for the trefoil $3_1=[2,3]=[3,2]$ on the exceptional line is:
\be\label{41}
P^{[23]}_{Adj, Exc}(q|A) 
= A^{12}\Big((1+q^{-4} + A + (q^3+q^{-3})(q A^2-q^{-1}A^5) + q^2A^3
- q^{-2}A^{15}-\frac{1-q^2+2q^4}{q^4}A^6-\\-\frac{1+q^2-q^4+q^6}{q^2}A^7-\frac{1-q^2+q^4+q^6}{q^2}
(A^8-q^{-2}A^{11})
-\frac{(q^2+q-1)(q^2-q-1)}{q^2}A^{12}+\\+\frac{1-q^4+q^6}{q^2}A^{13}
+ \{q\} \Big(q^{-1}A^4-[2]A^9+qA^{14}-qA^{16}-q^{-1}A^{17}+qA^{18}\Big)+\{q\}^2A^{10}\Big)
\ee
where $A=q^{N+2}$. The primary differential expansion in these variables is
\be
P^{[23]}_{Adj, Exc}(q|A)\ -\ 1 \ \ \  \vdots \ \ \  A^2(A^2+1+A^{-2})(A-1)^2(A+1)
\ee

\bigskip

\subsection{Universal description
\label{unidesc}}

Above formulas are fairly complicated, and the best way to look at them
is from the {\it universal} point of view.
In fact, most of them can be obtained by  substitution into the some universal expressions
of the particular values
of three (projective, universal, Vogel's) parameters $\alpha,\beta,\gamma$, relevant up to rescaling and permutations, according to the following Vogel's Table \cite{V2}:

\be
\begin{array}{|c|c|c|c|}
\hline   
\text{algebra} & \alpha & \beta & \gamma  \\
 \hline  
SU(N) &    -2 & 2 & N \\
 \hline  
SO(N) & -2 & 4 & N-4 \\
 \hline
Sp(N)  & -2 & 1 & \hbox{\small {$\frac{1}{2}$}}N+2\\
 \hline  
Exc(N) & -2 & N+4 & 2N+4 \\
 \hline
\end{array}
\label{VogelT}
\ee
where all exceptional simple Lie algebras belong to the $Exc$ line at special values of parameter:

\be
\begin{array}{|c||c|c|c|c|c|c|c|c|c|c|c|c|}
\hline 
N & -1  & -2/3 &0 & 1 & 2 & 4 & 8 \\
\hline  
Exc(N) & A_2 & G_2& D_4 & F_4 & E_6& E_7 & E_8 \\
\hline
\end{array}
\label{VogelExc}
\ee

\bigskip

Tables (\ref{VogelExc}) and (\ref{VogelT}) are derived from the following main observation of Vogel. Consider simple Lie algebra (extended by the automorphisms of its Dynkin diagram) with second Casimir's eigenvalue $2t$, in some arbitrary normalization . Then  symmetric square of adjoint decomposes in a uniform way, for all algebras:

\be \label{s2ad}
{\cal S}^2 Adj = 1+Y_2(\alpha)+Y_2(\beta)+Y_2(\gamma)
\ee
where $Y_2(\alpha), Y_2(\beta), Y_2(\gamma)$ have eigenvalues of the same Casimir operator
$4t-2\alpha, 4t-2\beta, 4t-2\gamma$, respectively.
One can show that $t  =  \alpha + \beta + \gamma $.
Actually this is definition of these parameters,
which evidently fix them  up to common multiplier and permutations.
Correspondingly they span the so-called Vogel's plane,
which is factor of projective plane over symmetric group, $P^2/S_3$, \cite{LM}.

Vogel \cite{Vogel,V2} gave a universal expressions for  dimension  of algebra:

\be \label{f3}
dim \, Adj =  \frac{(\alpha-2t)(\beta-2t)(\gamma-2t)}{\alpha \beta \gamma}
\ee
and for $Y_2(.)$ representations:

 \be
 \label{Y}
 dim \, Y_2(\alpha)=- \frac{\left( 3\,\alpha - 2\,t \right) \,\left( \beta - 2\,t \right) \,\left( \gamma - 2\,t \right) \,t\,\left( \beta + t \right) \,
      \left( \gamma + t \right) }{\alpha^2\,\left( \alpha - \beta \right) \,\beta\,\left( \alpha - \gamma \right) \,\gamma}
 \ee
These expressions for dimensions are universal in a sense that they are given by smooth (rational, in this case) functions of parameters, which at values from Vogel's Table (\ref{VogelT}), (\ref{VogelExc}) give dimensions for corresponding simple Lie algebra.
One can easily check that dimensions of two sides of (\ref{s2ad}) coincide at an arbitrary values of parameters.

Vogel \cite{Vogel,V2} and Landsberg and Manivel \cite{LM} have found a lot of universal formulas for dimensions of irreps of simple Lie algebras, belonging to powers of adjoint representation.
Quantization of most of them is already carried on. The universal character of adjoint representation (i.e. character of  adjoint, restricted on Weyl line, or, in other words, quantum version of (\ref{f3})) is given in \cite{MV12}:
\be
\chi_{Adj}(x\rho)\equiv f(x)=r+\sum_{\mu\in \Delta}e^{x(\mu,\rho)}=\prod _{\mu \in \Delta_+} \frac{e^{\frac{x}{2}(\theta+\rho,\mu)}-e^{-\frac{x}{2}(\theta+\rho,\mu)}}{e^{\frac{x}{2}(\rho,\mu)}-e^{-\frac{x}{2}(\rho,\mu)}} \\
f(x)=\frac{\sinh(x\frac{\alpha-2t}{4})}{\sinh(x\frac{\alpha}{4})}\frac{\sinh(x\frac{\beta-2t}{4})}{\sinh(x\frac{\beta}{4})}\frac{\sinh(x\frac{\gamma-2t}{4})}{\sinh(x\frac{\gamma}{4})}
\ee
where $r$ is the rank of the algebra, $\Delta (\Delta_+)$ is the set of all (all positive) roots, $\theta=\Lambda_{Adj}$ is the highest root.

Introducing, in agreement with (\ref{Weyl}),
\be
q=e^{\frac{x}{2}}
\ee
we get finally in the convenient form:

\be
{\cal D}_{Adj} = f(x)= -\frac{\{\sqrt{u}vw\}\{\sqrt{v}uw\}\{\sqrt{w}uv\}}{\{\sqrt{u}\}\{\sqrt{v}\}\{\sqrt{w}\}}
\ee

Note that for Chern-Simons theory this is Wilson average for unknot provided we take $x=2\pi /\delta, q=exp (\pi/\delta), \delta=\kappa + t$, where $\kappa$ is coupling constant in front of Chern-Simons action.  Note also that now theory is invariant w.r.t. the simultaneous rescaling of all 4 parameters $\alpha, \beta, \gamma, \kappa$, and quantization of coupling $\kappa$ means that it should be (an arbitrary) integer in the so-called minimal normalization, given in Table (\ref{VogelT}), (\ref{VogelExc}).

Quantum dimensions of $Y_2(.)$ are given in \cite{RMD}:

\be
{\cal D}_{Y_2}(\alpha) =
\frac{\{uvw\}\{u\sqrt{v}w\}\{uv\sqrt{w}\}\{v\sqrt{uw}\}\{w\sqrt{uv}\}\{vw/\sqrt{u}\}}
{\{\sqrt{u}\}\{u \}\{\sqrt{v}\}\{\sqrt{w}\}\{\sqrt{u/v}\}\{\sqrt{u/w}\}}
\ee

One can check that quantum dimensions of both sides of (\ref{s2ad}) coincide \cite{RMD}:

\be
{\cal S}^2 {\cal D}_{Adj} = \frac{{\cal D}_{Adj}^2(q) +{\cal D}_{Adj}(q^2)}{2}
= 1+D_{Y_2}(\alpha)+D_{Y_2}(\beta)+D_{Y_2}(\gamma)
\ee

It is a hypothesis of Deligne \cite{D13} (or universal characters' hypothesis), completely checked for $SL$ line \cite{D13} that quantum dimensions satisfy this and other standard characters' relations not only for points of Lie algebras, but also on the entire Vogel's plane. For the remaining antisymmetric subspace of square of adjoint the decomposition and dimensions are

\be
{\Lambda}^2 Adj = Adj + X_2
\ee

\be
dim(X_2) = \frac{(2t-\alpha)(2t-\beta)(2t-\gamma)(t+\alpha)(t+\beta)(t+\gamma)}{\alpha^2\beta^2\gamma^2}
\ee

and quantum dimensions satisfy
\be
{\Lambda}^2 {\cal D}_{Adj} = {\cal D}_{Adj} +  \frac{{\cal D}_{Adj}^2(q) -{\cal D}_{Adj}(q^2)
-2{\cal D}_{Adj}(q)}{2} = {\cal D}_{Adj} + {\cal D}_{X_2}
\ee
with \cite{D13}
\be
{\cal D}_{X_2} = {\cal D}_{Adj}\cdot\frac{\{u\sqrt{vw}\}\{v\sqrt{uw}\}\{w\sqrt{uv}\}}{\{u\}\{v\}\{w\}}
\left(\sqrt{uv}+\frac{1}{\sqrt{uv}}\right)\left(\sqrt{vw}+\frac{1}{\sqrt{vw}}\right)
\left(\sqrt{uw}+\frac{1}{\sqrt{uw}}\right)
\label{dimssquare}
\ee
The (q-powers of the half of the) corresponding universal expressions for the quadratic Casimir operators are \cite{Vogel}:
\be
\lambda_{Adj} = q^{t}= uvw, \ \ \ \ \
\lambda_{Y_2}(\alpha) = q^{2t-\alpha} = uv^2w^2, \ \ \ \ \
\lambda_{X_2} = q^{2t} = (uvw)^2
\label{evssquare}
\ee
We use here the notation from \cite{V2} for particular descendants of the adjoint representation.
Formulas for $Y_2(\beta)$ and $Y_2(\gamma)$ are obtained from $Y_2(\alpha)$
by cyclic permutations of $u,v,w$.

In the next Section we shall present similar formulas for decomposition of the cube of adjoint representation.

\section{The universal form of Rosso-Jones formula for 2 and 3 strands.
\label{URJ2st}}

Our aim in this section is to rewrite  Rosso-Jones expressions (\ref{RJ2}), (\ref{RJ}) for invariants of the torus knots/links in the universal form.
Its only group-depending elements are eigenvalues of second Casimir and quantum dimensions of representations Q, so we need  universal expressions for them.

\subsection{2-strand knots and links}

In this case $m=2$ and both Casimirs and quantum dimensions of irreps in decomposition of the square of adjoint are universal, as recalled in previous sections.
There are two diagrams Y with two boxes, symmetric square - i.e. a row of two boxes, and antisymmetric, a column of two boxes. For $n$ odd corresponding characters  $\varphi_{_Y} (\bar{\sigma}^{[2,n]})$ ,  evaluated on the only non-trivial element of $S_2$, are 1 and -1, respectively.

This allows one to rewrite (\ref{RJ2}), (\ref{RJ}) in the universal form for the 2-strand torus knots, i.e. $n=2k+1$:
\be
\boxed{
{\cal U}_{Adj}^{[2,n=2k+1]} = \frac{(uvw)^{4n}}{{\cal D}_{Adj}}\Big(1
+\lambda_{Y_2}(\alpha)^{-n}{\cal D}_{Y_2}(\alpha)
+ \lambda_{Y_2}(\beta)^{-n}{\cal D}_{Y_2}(\beta)
+\lambda_{Y_2}(\gamma)^{-n}{\cal D}_{Y_2}(\gamma)
- \lambda_{Adj}^{-n}{\cal D}_{Adj} - \lambda_{X_2}^{-n}{\cal D}_{X_2}\Big)
}
\label{URJ2str}
\ee

or

\be
{\cal U}^{[2,n]}_{\rm Adj} = \frac{(uvw)^{2n}}{{\cal D}_{Adj}}\cdot\Big((uvw)^{2n}- {\cal D}_{X_2}
+ u^{n} {\cal D}_{Y_2(\alpha)} + v^{n} {\cal D}_{Y_2(\beta)} + w^{n} {\cal D}_{Y_2(\gamma)}
-(uvw)^n{\cal D}_{\rm Adj}
\Big)
\ee

For $n=1$, i.e. for the unknot, ${\cal U}_{Adj}^{[2,1]}=1$,
for  general odd positive $n$  this is a polynomial
in positive powers of $u,v,w$, proportional to $(uvw)^{2n-2}$.

It is almost evident that this expression reproduces the 2-strand adjoint Kauffman polynomials
at the $SO(N)$ line $u=q^{-2}, \ v=q^4,\ w=A/q^3$, as long as dimensions and eigenvalues are reproduced by above universal expressions.

One may ask what it means at the $SU(N)$ line $u=q^{-2}, \ v=q^2,\ w=A$,
where the three non-unit dimensions in the symmetric square are
\be
D_{Y_2}(\alpha)=\frac{\{A\}^2\{Aq^3\}\{A/q\}}{[2]^2\{q\}^4}, \ \ \ \ \
D_{Y_2}(\beta)=\frac{\{A\}\{Aq\}\{A/q^3\}}{[2]^2\{q\}^4},\ \ \ \ \
D_{Y_2}(\gamma)=\frac{\{Aq\}\{A/q\}}{\{q\}^2}=D_{Adj}
\label{DSUY2}
\ee
and the coinciding
dimensions of two mutually-conjugate non-adjoint representations ($[31^{N-3}]$ and $[332^{N-3}]$, which form $X_2$ in the $SU(N)$ case, see (\ref{30}))
in the antisymmetric square are
\be
\frac{1}{2}{\cal D}_{X_2} = \frac{\{Aq^2\}\{Aq\}\{A/q\}\{A/q^2\}}{[2]^2\{q\}^4}
\label{DSUX2}
\ee
Clearly,
{ (\ref{URJ2str}) in this case is the uniform HOMFLY polynomial ${\cal H}_{\rm Adj}$}
from (\ref{uniH}), (\ref{30}).

To universally describe the 2-strand torus links, i.e. for $n=2k$, one has to change signs in
front of the two last items from minus to plus, since now we are evaluating characters of one-dimensional representations on the identity element of $S_2$, so they both are equal to 1:

\be
\boxed{
{\cal U}_{Adj}^{[2,n=2k]} = \frac{(uvw)^{4n}}{{\cal D}_{Adj}}\Big(1
+\lambda_{Y_2}(\alpha)^{-n}{\cal D}_{Y_2}(\alpha)
+ \lambda_{Y_2}(\beta)^{-n}{\cal D}_{Y_2}(\beta)
+\lambda_{Y_2}(\gamma)^{-n}{\cal D}_{Y_2}(\gamma)
+\lambda_{Adj}^{-n}{\cal D}_{Adj} +\lambda_{X_2}^{-n}{\cal D}_{X_2}\Big)
}
\label{URJ2strl}
\ee

The Rosso-Jones expression and its universalization closely resembles Okubo's formula \cite{Okubo} for eigenvalues of higher order Casimirs, used in \cite{MSV} to obtain universal expression for generating function of eigenvalues of higher Casimir operators.
\bigskip

\subsection{3-strand knots and links}

The universalization of general Rosso-Jones formula (\ref{RJ2}), (\ref{RJ}) for 3-strand knots/links is also possible. The cube of adjoint representation is decomposed as follows:
\be\label{decocube}
Adj^{\otimes 3} = {\cal S}^3(Adj)+2{S}_{[21]}(Adj)+{\Lambda}^3(Adj)
\ee
where the three terms in the sum correspond to the three components of the cube with different Young diagram symmetries, according to notations. First and third terms are 1-dimensional representations of the symmetric group $S_3$, second term is two-dimensional standards representation of $S_3$. Decompositions of these terms into universal irreps are, according to Theorem 3.8 of \cite{V2}:

\be
{\cal S}^3(Adj)=2X_1\oplus X_2 \oplus B(\alpha)\oplus B(\beta)\oplus B(\gamma)
\oplus Y_3(\alpha)\oplus Y_3(\beta)\oplus Y_3(\gamma)   \\
{\Lambda}^3(Adj)=X_0\oplus X_2 \oplus Y_2(\alpha) \oplus Y_2(\beta)\oplus Y_2(\gamma)
\oplus X_3(\alpha) \oplus X_3(\beta)\oplus X_3(\gamma)
\ee

\be
{S}_{[21]}(Adj)=2X_1\oplus 2X_2 \oplus Y_2(\alpha) \oplus Y_2(\beta)\oplus Y_2(\gamma)
\oplus B(\alpha)\oplus B(\beta)\oplus B(\gamma)
\oplus C(\alpha)\oplus C(\beta)\oplus C(\gamma)
\ee

$X_0$ is a singlet with unit dimension and eigenvalue, $X_1=Adj$,
representations $X_2$ and $Y_2$ are the same as appeared in the square of adjoint in sec.\ref{unidesc},
their dimensions and associated eigenvalues are given in (\ref{dimssquare}) and (\ref{evssquare}).

The plethysm (Adams rule) together with Deligne hypothesis gives the quantum dimensions of the three sectors:
\be
{\cal D}_{{\cal S}^3(Adj)}(q) = \frac{
{\cal D}_{Adj}(q)^3 + 3{\cal D}_{Adj}(q^2){\cal D}_{Adj}(q)+2{\cal D}_{Adj}(q^3)}{6},   \\
{\cal D}_{\Lambda^3(Adj)}(q) = \frac{
{\cal D}_{Adj}(q)^3 - 3{\cal D}_{Adj}(q^2){\cal D}_{Adj}(q)+2{\cal D}_{Adj}(q^3)}{6},  \\
{\cal D}_{S_{[21]}(Adj)}(q) = \frac{
{\cal D}_{Adj}(q)^3  - {\cal D}_{Adj}(q^3)}{3}
\label{Adthreedims}
\ee
where the plethystic replace $q\to q^n$ in the universal terms means $u\to u^n$, $v\to v^n$, $w\to w^n$.

To get the universal Rosso-Jones formula for 3-strand knots we note that characters' multipliers  $\varphi_{_Y} (\bar{\sigma}^{[3,n=3k \pm 1]})$ for fully symmetric and antisymmetric representations of $S_3$ are 1, and that for [21] contribution (standard two-dimensional representation of $S_3$) is (-1), see, e.g. \cite{S3}  . These values also  follows from the Adams rule (plethysm):

\be
\hbox{3-Adams}={\cal S}^3(Adj)-{S}_{[21]}(Adj)+{\Lambda}^3(Adj)
\ee

So:
\be
{\cal U}^{[3,n]}_{\rm Adj} = \frac{(uvw)^{6n}}{{\cal D}_{Adj}} \sum_{I} \pm\lambda_I^{-2n/3}D_I
\label{URJrstr1}
\ee

where the sign is plus for representations from the $3$ and $111$ sectors in $Adj^{\otimes 3}$,
and minus for those from the (one) $21$ sector.

This means that there are numerous cancellations and contributing to the sum
over $I$ in (\ref{URJrstr1})
are actually $10$ representations: seven
\be
X_0\oplus X_3(\alpha) \oplus X_3(\beta)\oplus X_3(\gamma)\oplus Y_3(\alpha)\oplus Y_3(\beta)\oplus Y_3(\gamma)
\ee
with the sign plus and three
\be
C(\alpha)\oplus C(\beta)\oplus C(\gamma)
\ee
with the sign minus.

According to Theorem 3.8 in \cite{V2} the corresponding Casimir's eigenvalues lead to
\be\label{vark}
\lambda_{X_i} = (uvw)^i,  \ i=0,1,2,3, \ \ \ \
\lambda_{Y_2}(\alpha) = uv^2w^2, \ \ \ \ \lambda_{Y_3}(\alpha) = v^3w^3,  \\
 \lambda_B(\alpha) = u^3v^2w^2, \ \ \ \ \lambda_C(\alpha) = u^{3/2}v^3w^3
\ee
Note that contributing to (\ref{URJrstr1}) are only representations, whose
eigenvalues are cubic in parameters $u,v,w$, so that $\lambda_I^{2/3}$ are
integer powers.

Starting from the cube of adjoint, the quantum dimensions are rarely known
in the universal form. A part of the problem is that even at the classical level
some of them are not just rational functions of parameters $\alpha,\beta,\gamma$,
but belong to certain extensions of this field (cubic in the case of $Adj^{\otimes 3}$), as shown in \cite{V2}.
However, it is very interesting and important that the Rosso-Jones formula appears insensitive to this complication!

The classical dimensions for {\it all} components of $Adj^{\otimes 3}$ are provided in the same theorem 3.8,
and quantization for {\it some} of them (for those, belonging to the {\it symmetric} cube)
is suggested in \cite{RMD}.
The classical dimension of $X_3$ is expressed through the parameters $\alpha$, $\beta$ and $\gamma$
by algebraic functions: it involves roots of a cubic equation with
coefficients made from these parameters, which make problem for explicit quantization.
However, since all the three $X_3$ has coincident eigenvalues, only the sum of all the three
enters (\ref{URJrstr1}), i.e. the character hypothesis eliminates this problem.

More exactly, the sum can be obtained by subtraction of known dimensions from that
of the antisymmetric cube:
\be
{\cal D}_{X_3} \equiv
{\cal D}_{X_3(\alpha)}+{\cal D}_{X_3(\beta)}+{\cal D}_{X_3(\gamma)} =
{\cal D}_{\Lambda^3(Adj)} - 1 - {\cal D}_{X_2} - {\cal D}_{Y_2(\alpha)} -
{\cal D}_{Y_2(\beta)} -{\cal D}_{Y_2(\gamma)}
\label{dimUX3}
\ee
The resulting explicit formula is long and not very informative,
therefore we do not present it here.

The essentially  new quantum dimensions at the cube level are \cite{RMD}
\be\label{dimBY3}
{\cal D}_{Y_3(\alpha)} = -\frac{\{uvw\}\{v\sqrt{w}\}\{w\sqrt{v}\} \{v\sqrt{uw}\}\{w\sqrt{uv}\}
\{uv\sqrt{w}\}\{uw\sqrt{v}\}\{vw/u\sqrt{u}\}\{vw\sqrt{u}\}}
{\{\sqrt{u}\}\{\sqrt{v}\}\{\sqrt{w}\} \{u\}\{u\sqrt{u}\}\{\sqrt{v}/u\}\{\sqrt{w}/u\}\{\sqrt{u/v}\}\{\sqrt{u/w}\}}   \\
{\cal D}_{B(\alpha)} = -\frac{\{uvw\}\{v\sqrt{uw}\}\{w\sqrt{uv}\}\{uv\sqrt{w}\}\{uw\sqrt{v}\}\{vw\sqrt{u}\}
\{u\sqrt{v}\}\{u\sqrt{w}\}\{uv/\sqrt{w}\}\{uw/\sqrt{v}\}}{\{\sqrt{u}\}\{u\}\{\sqrt{v}\}^2\{\sqrt{w}\}^2
\{\sqrt{v}/w\}\{\sqrt{w}/v\}\{\sqrt{v/u}\}\{\sqrt{w/u}\}}
\ee
and, finally, applying characters' hypothesis to relation from Theorem 3.8:

\be
Adj \otimes Y_2(\alpha)=Adj+X_2+Y_2(\alpha)+Y_3(\alpha)+B(\beta)+B(\gamma)+C(\alpha)
\ee
we get quantum dimension of $C(\alpha)$:

\be\label{dimC}
{\cal D}_{C(\alpha)} =
-\frac{\{uvw\}\{vw\}\{v\sqrt{w}\}\{w\sqrt{v}\}\{u\sqrt{vw}\}\{uv\sqrt{w}\}\{uw\sqrt{v}\}\{v\sqrt{uw}\}\{w\sqrt{uv}\}}
{\{\sqrt{u}\}^2\{u^{3/2}\}\{\sqrt{v}\}\{\sqrt{w}\}\{\sqrt{u/v}\}\{\sqrt{u/w}\}\{\sqrt{u}/v\}\{\sqrt{u}/w\}}\cdot
  \\
\cdot\left(\sqrt{uv}+\frac{1}{\sqrt{uv}}\right)\left(\sqrt{uw}+\frac{1}{\sqrt{uw}}\right)
\left(\sqrt{\frac{u}{vw}}+\sqrt{\frac{vw}{u}}\right)
\ee
It is easy to check that together with (\ref{dimssquare}) and (\ref{Adthreedims}) these expressions are consistent
with decompositions (\ref{decocube}).

Substituting all this into (\ref{URJrstr1}), we extend (\ref{URJ2str}) to

\be
\!\!\!\!\!\!\!\!\!\!\!\!
\boxed{
{\cal U}^{[3,n=3k\pm 1]}_{\rm Adj} = \frac{(uvw)^{4n}}{{\cal D}_{Adj}}\cdot\Big((uvw)^{2n}+ {\cal D}_{X_3}
+ u^{2n} {\cal D}_{Y_3(\alpha)} + v^{2n} {\cal D}_{Y_3(\beta)} + w^{2n} {\cal D}_{Y_3(\gamma)}
-u^{n} {\cal D}_{C(\alpha)} - v^{n} {\cal D}_{C(\beta)} - w^{n} {\cal D}_{C(\gamma)}
\Big)}
\label{URJ3str}
\ee
One can directly check that ${\cal U}^{[3,2]}_{\rm Adj}={\cal U}^{[2,3]}_{\rm Adj}$, which is implied by the topological invariance of the knot polynomials.

On the $SO(N)$ line $u=q^{-2},\ v=q^4,\ w=q^{-3}A$
this reproduces the answer of \cite{Ste} for Kauffman polynomial,
\be
{\cal K}_{\rm Adj}^{[3,n]} =  {K^{[3,n]}_{[11]}}  =
\frac{q^{-6n}A^{6n}}{d_{[11]}}\Big(q^{-2n}A^{-2n}d_{[333]} - A^{-2n}d_{[321]}+q^{2n}A^{-2n}d_{[3111]}
+q^{2n}A^{-2n}d_{[222]} -  \\
- q^{6n}A^{-2n}d_{[21111]}+q^{10n}A^{-2n}d_{111111}+1\Big)
\label{K3str}
\ee
with
\be
{\cal D}_{X_3} = d_{[222]}+d_{[3111]},   \\
{\cal D}_{Y_3(\alpha)} = d_{[333]}, \ \ \ \
{\cal D}_{Y_3(\beta)}= d_{[111111]}, \ \ \ \
{\cal D}_{Y_3(\gamma)} = 0,   \\
{\cal D}_{C(\alpha)} = d_{[321]}, \ \ \ \
{\cal D}_{C(\beta)} = d_{[211]}, \ \ \ \
{\cal D}_{C(\gamma)} = 0
\ee
Dimensions $Y_3(\gamma)$ and $C(\gamma)$ vanish for $SO(N)$  already at the classical level, since
they are proportional to $2\alpha+\beta \ \stackrel{SO(N)}{=}\ 0$.

\bigskip

One the $SU(N)$ line $u=q^{-2},\ v=q^2,\ w=A$, one reproduces the uniform HOMFLY provided
\be
{\cal D}_{Y_3(\alpha)} =\frac{\{Aq^5\}\{Aq\}^2\{A\}^2\{A/q\}}{\{q^3\}^2\{q^2\}^2\{q\}^2}=D_{[63^{N-2}]}, \ \ \ \ \
{\cal D}_{Y_3(\beta)} =\frac{\{A/q^5\}\{A/q\}^2\{A\}^2\{Aq\}}{\{q^3\}^2\{q^2\}^2\{q\}^2}=D_{[2221^{N-6}]},
\ee
$$
{\cal D}_{Y_3(\gamma)} = 1=D_{[0]},\ \ \ \ \
{\cal D}_{C(\alpha)} = 2\times \frac{\{Aq^4\}\{Aq^2\}\{A\}^2\{A/q\}\{A/q^2\}}{\{q^3\}^2\{q^2\}\{q\}^3}
=D_{[52^{N-3}1]}+D_{[543^{N-3}]},
$$
$$
{\cal D}_{C(\beta)} = 2\times \frac{\{A/q^4\}\{A/q^2\}\{A\}^2\{Aq\}\{Aq^2\}}{\{q^3\}^2\{q^2\}\{q\}^3}
=D_{[321^{N-5}]}+D_{[3332^{N-5}1]},\ \ \ \ \
{\cal D}_{C(\gamma)} = 0
$$
and
\be

{\cal D}_{X_3} = \frac{\{Aq^3\}\{Aq\}\{A/q\}\{A/q^3\}}{\{q^3\}^2\{q^2\}^2\{q\}^2}\cdot
\Big((q^2+4+q^{-2})(A^2+A^{-2})-(3q^4+2q^2+2+2q^{-2}+3q^{-4})\Big)
\ee
Again, for $SU(N)$ already the classical dimension ${\cal D}_{C(\gamma)} \ \sim\
\alpha+\beta \stackrel{SU(N)}{=} 0$.
We remind that $X_3$ is actually a sum of three representations,
namely, for $SU(N)$
\be
{\cal D}_{X_3} = {\cal D}_{X_3(\alpha)}+{\cal D}_{X_3(\beta)}+{\cal D}_{X_3(\gamma)} 
=2\times \frac{\{Aq^3\}\{Aq^2\}\{Aq\}\{A/q\}\{A/q^2\}\{A/q^3\}}{\{q^3\}^2\{q^2\}^2\{q\}^2}
+ \frac{\{Aq^3\}\{Aq\}^2\{A/q\}^2\{A/q^3\}}{\{q^3\}^2\{q\}^4} =  \\
=D_{[432^{N-4}1]} + D_{[41^{N-4}]} + D_{[4443^{N-4}]}
\ee
in excellent accordance with (\ref{dimUX3}).
The three dimensions from $Adj^{\otimes 3}$, which do not contribute to
the Rosso-Jones formula (\ref{URJ3str}), for $SU(N)$ are:
\be
{\cal D}_{B(\alpha)} = \frac{\{Aq\}\{A\}^2\{A/q^3\}}{\{q\}^2\{q^2\}^2}=D_{[2^21^{N-4}]}, \ \ \ \ \
{\cal D}_{B(\beta)} = \frac{\{Aq^3\}\{A\}^2\{A/q\}}{\{q\}^2\{q^2\}^2}=D_{[42^{N-2}]}, \\
{\cal D}_{B(\gamma)} = \frac{\{Aq^3\}\{Aq\}^2\{A/q\}^2\{A/q^3\}}{\{q\}^4\{q^3\}^2}=D_{[432^{N-4}1]}
\ee

\bigskip

The extension of the formula (\ref{URJ3str}) to the case of 3-strand links is immediate but looks much longer. In this case all the terms in the expansion (\ref{decocube}) contribute, since now we evaluate characters, $\varphi_{_Y} (\bar{\sigma}^{[3,n=3k]})$,  on identity element of group, which gives dimensions of representations, i.e. 1,1,2 for symmetric, antisymmetric and [21] cases, respectively.

So, for links we get

\fr{
{\cal U}^{[3,n=3k]}_{\rm Adj} = \frac{(uvw)^{12k}}{{\cal D}_{Adj}}\cdot\Big((uvw)^{6k}+6(uvw)^{4k}{\cal D}_{X_1} +6(uvw)^{2k}{\cal D}_{X_2}+{\cal D}_{X_3}+3u^{2k}(uvw)^{2k}{\cal D}_{Y_2(\alpha)}+\\+3v^{2k}(uvw)^{2k}{\cal D}_{Y_2(\beta)} +3w^{2k}(uvw)^{2k}{\cal D}_{Y_2(\gamma)}+3(vw)^{2k}{\cal D}_{B(\alpha)}+3(uw)^{2k}{\cal D}_{B(\beta)}+3(uv)^{2k}{\cal D}_{B(\gamma)}+\\+2u^{3k} {\cal D}_{C(\alpha)}+2v^{3k} {\cal D}_{C(\beta)} + 2w^{3k} {\cal D}_{C(\gamma)}
+ u^{6k} {\cal D}_{Y_3(\alpha)} + v^{6k} {\cal D}_{Y_3(\beta)} + w^{6k} {\cal D}_{Y_3(\gamma)}
\Big)
\label{URJ3strl}
}

\bigskip

An extension to four and more strands is more difficult, because much less is known
about universal formulas for dimensions (even classical) in higher powers of adjoint representation.
However, as we discover above in the 3-strand example, the most difficult questions
like quantization of individual dimensions ${\cal D}_{X_3(\alpha)}$,
which even classically are algebraic functions of the universal parameters,
can be irrelevant for knot theory, at least for the torus knots, so one can hope that similar phenomena can happen in higher powers, also.

\bigskip

Non-torus knots is another challenge.
As we will now demonstrate with the example of the simplest non-torus $4_1$,
their knot polynomials can also be lifted to the universal level,
though a systematic way to do so still needs to be developed.

\section{Universal knot polynomial for the figure-eight knot $4_1$
\label{twist}}

The uniform adjoint HOMFLY is equal to
\be\label{uniHfe}
{\cal H}^{4_1}_{Adj}(q|A)\ =\ 1 \  + \
\Big(q^2+q^{-2} + \{q^2\}^2\Big) \cdot\frac{q^3+q^{-3}}{q+q^{-1}}\cdot\{A\}^2
\  +\ \left(\frac{q^3+q^{-3}}{q+q^{-1}}\cdot\{A\}^2\right)^2 ,
\ee
while the adjoint Kauffman polynomial for $4_1$ was found in \cite{NawataKAdj}:

\be
{\cal K}^{4_1}_{Adj} = K^{4_1}_{[11]} =A^{4}\Big(1-q^{-2}+2q^{-6}-q^{-8}-q^{-10}+q^{-12}\Big)+
A^{3}\Big(-q^3+2q-4q^{-3}+3q^{-5}+2q^{-7}-3q^{-9}+q^{-13}\Big)+ \\
+A^{2}\Big(q^6-3q^4+6-6q^{-2}-4q^{-4}+7q^{-6}-q^{-8}-3q^{-10}+q^{-12}\Big)+
A\Big(2q^7-q^5-6q^3+7q+4q^{-1}-10q^{-3}+2q^{-5}+ \\ +5q^{-7}-2q^{-9}-q^{-11}\Big)-
\Big(q^{10}-q^8-5q^6+7q^4+3q^2-13+3q^{-2}+7q^{-4}-5q^{-6}-q^{-8}+q^{-10}\Big)+ \\ +
A^{-1}\Big(-q^{11}-2q^9+5q^7+2q^5-10q^3+4q+7q^{-1}-6q^{-3}-q^{-5}+2q^{-7}\Big)+
A^{-2}\Big(q^{12}-3q^{10}-q^8+7q^6-4q^4- \\ -6q^2+6-3q^{-4}+q^{-6}\Big)+A^{-3}\Big(q^{13}-3q^9+2q^7+3q^5-4q^3+2q^{-1}-q^{-3}\Big) +
A^{-4}\Big(q^{12}-q^{10}-q^8+2q^6-q^2+1\Big)\nn
\ee

\bigskip

The universal knot polynomial is given by
\be
{\cal U}^{4_1}_{Adj} = ((444))-((443))+((442))+((433)) - ((432))-2((333))+((322))-2((321))+((320))
+ 2((311))-   \\
-6((222))+6((221)-2((220))-4((211))+((210))+((200))+2((111))+2((110))-6((100))+11-  \\
\!\!\!\!\!\!\!\!\!\!\!\!\!\!\!\!\!\!\!\!\!\!\!
-((2,0,-1))+2((1,0,-1))-2((1,1,-1))
\label{Ua41}
\ee
Since $4_1$ is fully amphicheiral, the polynomial is symmetric under the change
\be
(u,v,w)\longrightarrow (u^{-1},v^{-1},w^{-1})
\label{mirr}
\ee
because of this in (\ref{Ua41}) we use the notation
\be
((444))=(uvw)^4+(uvw)^{-4},   \\
((300)) = u^3+v^3+ w^3+u^{-3}+v^{-3}+w^{-3},
((321)) = u^3v^2w+u^{-3}v^{-2}w^{-1} + \text{5 permutations},
  \\
((2,0,-1)) = u^2/v+v/u^2+u^2/w+w/u^2+v^2/u+u/v^2+v^2/w+w/v^2+w^2/u+u/w^2+w^2/v+v/w^2,   \\
((1,0,-1))=u/v+u/w+v/w+v/u+w/u+w/v,   \\
\ldots
\ee
(note non-trivial multiplicities).
Expression (\ref{Ua41}) reproduces the uniform HOMFLY and Kauffman polynomials
at the $SU(N)$ and $SO(N)$ lines $uv=1$ and $u^2v=1$.
The ambiguity left by comparison with uniform HOMFLY and Kauffman polynomials is proportional to
\be
\frac{(uv-1)(vw-1)(uw-1)(u^2v-1)(uv^2-1)(vw^2-1)(v^2w-1)(uw^2-1)(u^2w-1)}{(uvw)^4}
\ee
times a rational function, which is {\it odd} under (\ref{mirr}).
Adding such polynomials would increase the power of (\ref{Ua41}).

\bigskip

At the exceptional line $w=u^2v^2$
\be
{\cal U}^{4_1}_{Adj}-1 \ \sim \
(uv)^{-12}(1-uv)\Big(1-(uv)^2\Big)\Big(1+(uv)^2+(uv)^4\Big)\Big(1-u-v+3uv+\ldots + (uv)^{17}\Big)
\ee

\section{On properties of the universal polynomials
\label{prop}}

As mentioned in the Introduction,
our actual calculation started from the uniform HOMFLY and attempts to unify them
with the Kauffman polynomials into a universal expression.
It is instructive to present  some more details about these  polynomials
and their properties, which has a natural extension to the universal level.

Let us first list the properties that we are going to check for the universal polynomials of the concrete knots.
\begin{itemize}
\item[{\bf i)}] The special polynomial property:
\be
{\cal U}^{\cal K}_{Adj}(u=1,v=1,w)=\left[\sigma^{\cal K}_\Box (w)\right]^2
\ee
where $\sigma^{\cal K}_\Box (w)$ is a universal special polynomial in the fundamental representation (this implies that the universality is preserved even in the fundamental representation for the special polynomial).
\item[{\bf ii)}] The ``Alexander'' property:
\be
{\cal U}^{\cal K}_{Adj}(u,v,w)\Big|_{uvw=1}=1
\ee
The condition $uvw=1$ reduces to trivial Abelian factors for the concrete groups and is equivalent to $N=0$ in the $SU(N)$ case, hence, the name Alexander.
\item[{\bf iii)}] The differential expansion (related to the Alexander property above):
\be
{\cal U}^{\cal K}_{Adj}(u,v,w)-1\ \  \vdots\ \ (uvw-1)(uvw+1)
\ee
The remainder of this division can be further refined and depends on the knot.
\end{itemize}

Below we consider examples of concrete knots, manifest expressions for their polynomials and check properties (i)-(iii).

\subsection{Trefoil}

\subsubsection{The uniform HOMFLY polynomial}
For the trefoil the uniform HOMFLY (\ref{uniH}) is given by a remarkably simple expression:
\be
\boxed{
{\cal H}^{[2,3]}_{Adj}
= A^8\cdot \Big(A^{-2}(q^2+q^{-2})-q^2+1-q^{-2}\Big)^2\ +\ A^8\cdot(q-q^{-1})^2\cdot(A^2-A^{-2})
}
\label{uniH23}
\ee
Here $A=q^N$.
This formula certainly coincides with (\ref{30}) for $n=3$ and with (\ref{URJ2str}) at
the $SU(N)$ line $u=q^{-2},\ v=q^2,\ w=A$. Expressions (\ref{uniH23}) and (\ref{uniHfe}) are also in accord with that in \cite[p.25]{Morton}.

We immediately check that:

\paragraph{ii)} Its special polynomial is a square of the fundamental one:
\be
{\cal H}^{[2,3]}_{Adj}(q=1|A) = \Big((2-A^2)\cdot A^2\Big)^2 = \Big(H^{[2,3]}_{_\Box}(q=1|A)\Big)^2
\label{spepo31}
\ee
This is instead of
\be
H^{[2,3]}_{Adj(N)}(q=1|A) = \Big((2-A^2)\cdot A^2\Big)^N = \Big(H^{[2,3]}_{_\Box}(q=1|A)\Big)^N
\ee
for the usual adjoint colored HOMFLY at fixed $N$.

\paragraph{ii)} Its ``Alexander" polynomial is just unity:
\be
{\cal H}^{[2,3]}_{Adj}(q|A=1) = 1
\ee

\paragraph{iii)} Its differential expansion starts from $\{A\}^2$:
\be
{\cal H}^{[2,3]}_{Adj} -1 \ \ \vdots \ \ (A-A^{-1})^2
\label{diffexpanHa31}
\ee

\subsubsection{The adjoint Kauffman polynomial}

The Kauffman polynomial for the trefoil ${\cal K}^{[2,3]}_{Adj}(q,A)$ can be read off from (\ref{adjkau}) for $n=3$.
\paragraph{i).}
The special polynomials at $q=1$ are the same for $SU(N)$ and $SO(N)$, provided both are expressed
in terms of $A$ (which originally were identified respectively with $q^N$ and $q^{N-1}$):
\be
{K_R} (q=1) =  K_{[1]}^{|R|}(q=1) \ = \
 H_{[1]} ^{|R|}(q=1) =  {H_R} (q=1)
\ee
We already saw in (\ref{spepo31}) that  the uniform adjoint HOMFLY ${\cal H}_{Adj}$ at $q=1$
is also a square of the fundamental special polynomial, and now we see the reason: this is exactly the property of the adjoint Kauffman polynomial $K_{[11]}$,
\be
{K_{[11]}} (q=1)=H_{[1]}^2(q=1)
\ee
and if there is the universality, the same should be true for the uniform HOMFLY.

\paragraph{ii).}
Also
\be
\ \ \ {\cal K}^{[2,3]}_{Adj}(q|A=\pm q) =1,
\ee
and
\paragraph{iii).}
\be
\ \ \ {\cal K}^{[2,3]}_{Adj}(q|A)\ -\ 1 \ \ \vdots \ \ \ (Aq+1)\{A/q\}
\ee

\subsubsection{The universal polynomial}

It is not a surprise now that the universal polynomial,
which for the trefoil is explicitly equal to
\be\label{utr}
{\cal U}^{[2,3]}_{\rm adj} = 1 + \Big(1-(uvw)^2\Big)^2\cdot\Big(
-2S_{666}+3S_{665}-S_{555}-S_{444}+3S_{442}-S_{222}-1\Big) +   \\
+ \Big(1-(uvw)^2\Big)(1-uv)(1-uw)(1-vw)\cdot
\Big(S_{666}  +S_{555}-3S_{544} -S_{444}+3S_{433}+3S_{332} +S_{222}\Big) =   \\
  \\
=(uvw)^4\Big(-u^6v^6w^6+(u^6v^6w^5+v^6w^6u^5+w^6u^6v^5)-(u^6v^5w^5+v^6w^5u^5+w^6u^5v^5)- \\
-(u^5v^4w^4+v^5w^4u^4+w^5u^4v^4) \
+(u^5v^4w^3+v^5w^4u^3+w^5u^4v^3+u^5w^4v^3+v^5u^4w^3+w^5v^4u^3)+ \\
+3u^4v^4w^4 -(u^4v^4w^3+v^4w^4u^3+w^4u^4v^3)
+(u^4v^3w^3+v^4w^3u^3+w^4u^3v^3) -  \\
-(u^4v^2w^2+v^4w^2u^2+w^4u^2v^2)-(u^3v^3w^2+v^3w^3u^2+w^3u^3v^2)+(u^3v^2w^2+v^3w^2u^2+w^3u^2v^2)- \\
-(u^3v^2w+v^3w^2u+w^3u^2v+u^3w^2v+v^3u^2w+w^3v^2u)-  \\
-2u^2v^2w^2+(u^2+v^2+w^2)+(uv+vw+wu)+1\Big)
\ee

\paragraph{i)} satisfies the generalized factorization property at $u=v=1$
\be
{\cal U}^{[2,3]}_{Adj}(u=v=1,w) =\Big( w^2(w^2-2)\Big)^2
\label{spepoU23}
\ee
\paragraph{ii)}
it satisfies the Alexander identity
\be
{\cal U}^{[2,3]}_{Adj}(u,v,w)\Big|_{uvw=1} =1
\ee
and
\paragraph{iii)} there is a
differential expansion:
\be
{\cal U}^{[2,3]}_{Adj}(u,v,w ) -1 =
(uvw-1)(uvw+1)\Big(-S_{888}+3S_{887}  -3S_{877} -3S_{766} + 6S_{765}+
  \\
+2S_{666}  -3S_{644}-3S_{554} + S_{222} + 1\Big)
\label{diffexpanU23}
\ee
In these formulas $S_{abc}$ are the elementary symmetric polynomials,
\be
S_{abc} = \frac{1}{6}\Big( u^av^bw^c + 5\ \text{permutations}\Big)
\ee
Note that $S_{aaa} = (uvw)^a$, but $u^av^aw^b +v^aw^au^b+w^au^av^b = 3S_{aab}$.

There is also a couple of weaker properties:
at $w=1$
\be
{\cal U}^{[2,3]}_{Adj}(w=1,u,v)-1 = -(uv-1)^2(uv+1)\Big(u^6v^6-u^4v^4(u^2+v^2+u+v)
+u^4v^4+2u^3v^3+2u^2v^2+uv+1\Big)
\ee
and also at $w=-1$
\be
{\cal U}^{[2,3]}_{Adj}(w=1,u,v)-1 = -(uv-1)^2(uv+1)\Big(2u^7v^7+3u^6v^6 -  \\
-u^4v^4(2u^3v^2+2v^3u^2 +u^2-u+v^2-v)
+u^4v^4+2u^3v^3+2u^2v^2+uv+1\Big)
\ee

\subsubsection{Exceptional groups}

\paragraph{iii)}
For exceptional groups  we get an additional factorization in differential expansions:
$\frac{{\cal U}^{[2,3]}_{Adj}-1}{(uvw)^2-1}$ is divisible by

\be
\begin{array}{lc}
(1-q^2)(1+q^{2/3})&\text{for}\  G_2   \\
(1-q^2)(1+q^4)(1+q+q^2)&\text{for}\ F_4   \\
(1-q^6)(1+q^2)&\text{for}\ E_6   \\
(1-q^6)(1+q^2)(1+q^4)&\text{for}\ E_7   \\
(1-q^6)(1+q^2)(1+q^2+q^4+q^6)+q^8)=(1-q^{10})(1+q^2)(1+q^2+q^4)&\text{for}\ E_8
\end{array}
\ee

In general on the exceptional line $w=u^2v^2$ and from (\ref{utr}) one gets:
\be
{\cal U}^{[2,3]}_{Adj}(w=u^2v^2,u,v) -1 = -(uv-1)\Big((uv)^6-1\Big)\cdot
\Big((uv)^{23} -(uv)^{22}(u+v) + \ldots + 1\Big)
\label{uni23exc}
\ee
To compare, on the $SU$ and $SO/Sp$ lines $v=u^{-1}$ and $v=u^{-2}$
one respectively has
\be
{\cal U}^{[2,3]}_{Adj}(v=u^{-1},u,w) -1 = \frac{(w^2-1)^2}{u^2}\Big(
\cdot(u^3-2u^2 +u)w^6+(u^4-u^2+1)w^4-2u^2w^2-u^2\Big)
\ee
and
\be
{\cal U}^{[2,3]}_{Adj}(v=u^{-2},u,w) -1 = \frac{(w^2-u^2)(w+u^2)}{u^{11}}\Big((u-1)(u^2-1)w^7
+ \ldots + u^7\Big)
\ee

\subsubsection{Examples of exceptional knot polynomials for the trefoil}

We list here first adjoint exceptional knot polynomials, since they have never been calculated so far. We added for comparison the cases of small classical groups.

{\footnotesize
\be
P^{[3_1]}_{Adj, A_1}(q) = J^{[3_1]}_{[2]} =  q^4\Big(1+q^6-q^{10}+q^{12}-q^{14}-q^{16}+q^{18}\Big)
\nn \\
P^{[3_1]}_{Adj, A_2}(q) = H^{[3_1]}_{[21]}(q|A=q^3)=
q^8\Big(1+2q^4-2q^6+2q^8-2q^{10}+2q^{12}-4q^{14}+3q^{16}-2q^{18}+2q^{20}-2q^{22}+q^{24}\Big)
 \nn \\
P^{[3_1]}_{Adj, A_3}(q) = H^{[3_1]}_{[211]}(q|A=q^4)=
q^{12}\Big(1+2q^4-q^8+q^{10}-2q^{12}+q^{14}-q^{16}-2q^{18}+3q^{20}-2q^{22}+q^{24}+q^{26}-2q^{28}+q^{30}\Big)\nn\\
P^{[3_1]}_{Adj, A_4}(q) = H^{[3_1]}_{[2111]}(q|A=q^5)=
q^{16}\Big(1+2q^4+q^8-2q^{10}+q^{12}-2q^{14}+q^{16}-2q^{18}+q^{20}-2q^{22}+3q^{24}-2q^{26}+q^{28}
+q^{32}-2q^{34}+q^{36}\Big) \nn \\
\ldots \nn \\
P^{[3_1]}_{Adj, D_1}(q) = K^{[3_1]}_{[11]}(q|A=q)=1 \nn \\
P^{[3_1]}_{Adj, D_2}(q) = K^{[3_1]}_{[11]}(q|A=q^3)=q^4\Big(1+q^6-q^{10}+q^{12}-q^{14}-q^{16}+q^{18}\Big)
= P^{[3_1]}_{Adj, A_1}(q) \nn \\
P^{[3_1]}_{Adj, D_3}(q) = K^{[3_1]}_{[11]}(q|A=q^5)=
q^{12}\Big(1+2q^4-q^8+q^{10}-2q^{12}+q^{14}-q^{16}-2q^{18}+3q^{20}-2q^{22}+q^{24}+q^{26}-2q^{28}+q^{30}\Big)
= P^{[3_1]}_{Adj, A_3}(q)\nn\\
P^{[3_1]}_{Adj, D_4}(q) = K^{[3_1]}_{[11]}(q|A=q^7)=
q^{20}\Big(1+q^4+2q^6-2q^{10}+2q^{12}+q^{14}-4q^{16}-2q^{18}+3q^{20}-5q^{24}+q^{26}+\nn \\
+5q^{28}-q^{30}-3q^{32}+2q^{34}+2q^{36}-2q^{38}-q^{40}+q^{42}\Big)  \nn\\ \nn \\
P^{[3_1]}_{Adj, D_5}(q) = K^{[3_1]}_{[11]}(q|A=q^9)=
q^{28}\Big(1+q^4+q^6+q^8-2q^{20}-2q^{22}+q^{26}-q^{28}-q^{30}-q^{32}+q^{34}+3q^{36}-q^{40}-\nn\\
-q^{42}+q^{44}+q^{46}-q^{50}-q^{52}+q^{54}\Big) \nn \\
\ldots \nn \\
P^{[3_1]}_{Adj, G_2}(q) = q^{12}\Big(1+q^4+q^{14/3}+q^{16/3}-q^{20/3}-q^{22/3}-q^8+
q^{28/3}+2q^{10}+q^{32/3}-q^{34/3}-2q^{12}-\nn \\
-2q^{38/3}-q^{40/3}+q^{14}+2q^{44/3}+2q^{46/3}-2q^{50/3}
-3q^{52/3}-2q^{18}+2q^{58/3}+3q^{20}+q^{62/3}-2q^{22}-\nn \\
-q^{68/3}-q^{70/3}+2q^{74/3}
+2q^{76/3}-q^{80/3}-q^{82/3}-q^{28}+q^{30}\Big) \nn \\ \nn \\
P^{[3_1]}_{Adj, F_4}(q) = q^{32}\Big(1+q^4+q^7+q^8+q^{16}-q^{18}+q^{20}-q^{21}-2q^{22}-q^{23}-\nn\\
-q^{25}-q^{26}+q^{27}+q^{28}-q^{30}-2q^{34}-q^{35}+q^{36}+q^{37}-q^{38}+q^{39}+3q^{40}+q^{41}-\nn\\
-q^{42}-q^{45}-q^{46}+q^{48}+q^{52}+q^{53}-q^{54}-q^{55}-q^{58}+q^{60}\Big) \nn \\ \nn\\
P^{[3_1]}_{Adj, E_6}(q) = q^{44}\Big(1+q^4+q^8+q^{10}+q^{16}-q^{24}-2q^{28}-q^{30}-q^{32} - \nn \\
-q^{40}-q^{44}+q^{48}+3q^{52}-q^{58}+q^{68}-q^{72}-q^{76}+q^{78}\Big) \nn \\ \nn \\
P^{[3_1]}_{Adj, E_7}(q) = q^{68}\Big(1+q^4+q^{10}+q^{14}+q^{20}+q^{24}-q^{26}+q^{28}-q^{30}-2q^{36}+\nn\\
+q^{38}-2q^{40}-q^{44}-q^{46}+q^{48}-2q^{50}+q^{52}-q^{54}-q^{60}+q^{62}-2q^{64}+2q^{66}-q^{68}+\nn\\
+q^{70}+q^{72}-q^{74}+3q^{76}-q^{78}+q^{80}-q^{84}+q^{86}-q^{88}+q^{90}-q^{92}+q^{100}
-q^{102}+q^{104}-q^{106}-q^{112}+q^{114}\Big) \nn \\ \nn \\
P^{[3_1]}_{Adj, E_8}(q) = q^{116}\Big(1+q^4+q^{14}+q^{22}+q^{28}+q^{36}-q^{42}+q^{44}-q^{50}-\nn\\
-q^{56}-q^{60}+q^{62}-2q^{64}-q^{72}-q^{74}+q^{76}-q^{78}-q^{82}+q^{84}-q^{86}-q^{88}+q^{92}-\nn\\
-q^{96}+q^{102}-2q^{104}+q^{106}+q^{110}-q^{112}+q^{114}+q^{116}-q^{122}+3q^{124}-q^{126}+\nn\\
+q^{132}-q^{136}+q^{138}-q^{144}+q^{146}-q^{152}+q^{164}-q^{166}+q^{172}-q^{174}-q^{184}+q^{186}\Big)
\nn
\ee
}

\subsection{General 2-strand torus knots $[2,2k+1]$}

\subsubsection{The HOMFLY polynomial}
For the entire one-parametric family of the 2-strand knots, the HOMFLY polynomial in the fundamental representation is
\be
H^{[2,2k+1]}_{_\Box}= A^{2k+1}\cdot\frac{q^{-2k-1}\{Aq\}-q^{2k+1}\{A/q\}}{\{q^2\}}
\ee
and the special polynomial is equal to
\be
H_{[1]}^{[2,2k+1]}(q=1)=\Big(k+1-kA^2\Big)\cdot A^{2k}
\ee
The uniform HOMFLY polynomial in the adjoint representation, a generalization of (\ref{uniH23})
turns out to be
\be
{\cal H}^{[2,2k+1]}_{Adj}(q|A) = A^{4k+4}\Big(1-X_k^2\Big)+ A^{4k}\Big(A^2X_k-A\{A\}Y_k\Big)^2
+ A^{4k+4}\{A^2\}\{q\}^2\cdot Z_k
\label{adjuniH2str}
\ee
where
\be
X_k = \frac{q^{2k-1}+q^{-2k+1}}{q+q^{-1}}, \ \ \ \ \
Y_k =\frac{q^{2k+2}-q^{-2k-2}}{q^2-q^{-2}}
\ee
and
\be
Z_k = \sum_{i=0}^{k-1} A^{4(k-1-i)}\cdot Y_i\cdot \Big(Y_i+A^2Y_{i-1}\Big)
\ee
e.g.
\be
Z_1 = 1, \ \ \ Z_2 = A^4+(q^2+q^{-2})A^2+(q^2+q^{-2})^2,  \\
Z_3 = A^8+(q^2+q^{-2})A^6+(q^2+q^{-2})^2A^4+(q^2+q^{-2})(q^4+1+q^{-4})A^2+(q^4+1+q^{-4})^2,
\ \ \ \ldots
\ee
\paragraph{i)}
This uniform HOMFLY satisfies
\be
{\cal H}^{[2,2k+1]}_{Adj}(q=1|A) = \Big(H_{[1]}^{[2,2k+1]}(q=1|A)\Big)^2
\ee
and
\paragraph{ii)}
\be
{\cal H}^{[2,2k+1]}_{Adj}(q|A=1) =1,
\ee
moreover,
\paragraph{iii)}
the deviation from unity is always quadratically small:
\be
{\cal H}^{[2,2k+1]}_{Adj}(q|A)-1 \ \vdots\ \{A\}^2
\ee

\subsubsection{The Kauffman polynomial}
The normalized Kauffman polynomial for 2-strand knots is given by the Rosso-Jones formula
(\ref{adjkau}),
\be
{\cal K}^{[2,n]}_{Adj}= K^{[2,n]}_{[11]}=
\frac{q^{-4n}A^{4n}}{d_{[11]}}\left( A^{-2n}d_{[22]} - q^{2n}A^{-2n}d_{[211]}+q^{6n}A^{-2n}d_{[1111]}+
q^{-n}A^{-n}d_{[2]}-q^nA^{-n}d_{[11]}+1
\right)
\label{adjkau2str}
\ee
and satisfies
\paragraph{i)}
\be
{\cal K}^{[2,2k+1]}_{Adj}(q=1|A) = \Big(\sigma^{[2,2k+1]}(A)\Big)^2
\ee
and
\paragraph{ii)}
\be
{\cal K}^{[2,2k+1]}_{Adj}(q|A=\pm q) =1,
\ee
\paragraph{iii)} Its deviation from unity is only linear, but always has an additional factor $(Aq-1)$:
\be
{\cal K}^{[2,2k+1]}_{Adj}(q|A)\ -\ 1 \ \ \vdots \ \ \ (Aq+1)\{A/q\}
\ee
We remind that in this case $A=q^{N-1}$, so $A=q$ corresponds to the Abelian $SO(2)$ group.

\subsection{Other knots}

The same properties i)-iii) are also true for the 3-strand torus knots and also for the figure-eight knot. For this later, in particular,
\be
{\cal U}^{4_1}_{Adj}(u=v=1,w) = \Big(w^2-1+w^{-2}\Big)^2
\ee
and the properties i)-iii) for the $SU(N)$ line look like
\be
{\cal H}^{4_1}_{Adj}(q=1|A) = \Big(A^2-1+A^{-2}\Big)^2=
\Big(\sigma^{4_1}(A)\Big)^2,   \\
{\cal H}^{4_1}_{Adj}(q|A=1) =1,   \\
{\cal H}^{4_1}_{Adj}(q|A)\ -\ 1 \ \ \vdots \ \ \{A\}^2
\label{propetiesUH41}
\ee
moreover, the complete (refined) differential expansion ({\bf iii}) is in this case
\be
{\cal H}^{4_1}_{Adj}(q|A)\ =\ 1 \  + \
\Big(q^2+q^{-2} + \underline{\{q^2\}^2}\Big) \cdot\frac{q^3+q^{-3}}{q+q^{-1}}\cdot\{A\}^2
\  +\ \left(\frac{q^3+q^{-3}}{q+q^{-1}}\cdot\{A\}^2\right)^2
\ee
If not the underlined term, this differential expansion would very much resemble
that for the symmetric representations \cite{IMMMfe},
\be
{\cal H}^{4_1}_{[2]}(q|A) = 1 + (q+q^{-1})\{Aq^2\}\{A/q\} + \{Aq^3\}\{Aq^2\}\{A\}\{A/q\},   \\
{\cal H}^{4_1}_{[r]}(q|A) = 1 + \sum_{k=1}^r\frac{[r]!}{[k]![r-k]!}\prod_{i=0}^{k-1} \{Aq^{2+i}\}\{Aq^{i-1}\}
\ee
As to underlined $\{q\}^2$ term it first appeared in \cite{Ano21},
see also \cite{MMM21} for a little more about such terms.

Similarly, for $SO(N)$ line we have:
\be
{\cal K}^{4_1}_{Adj} - 1 \ \ \vdots \ \ (Aq+1)\cdot\{A/q\}
\ee

\section{Conclusion}

In this paper we constructed {\bf the universal adjoint knot polynomials for the 2-strand torus knots (formula (\ref{URJ2str})) and links (formula (\ref{URJ2strl})), for the 3-strand torus knots (formula (\ref{URJ3str})) and links (formula (\ref{URJ3strl})), see also formulas (\ref{vark}) for the eigenvalues of the cut-and-join operator, $\varkappa_R$ and formulas for the universal quantum dimensions:  (\ref{dimBY3}), (\ref{dimC}) and (\ref{dimUX3}) with (\ref{Adthreedims}), (\ref{DSUY2}) and (\ref{DSUX2})}.

We also proposed {\bf a universal adjoint knot polynomial for the figure eight knot in (\ref{Ua41})}.

These are the first examples of universal expressions for non-trivial knots, and they provide a strong evidence that all adjoint colored
knot polynomials exhibit Vogel's universality and can be lifted to entire
Vogel's plane, so that the corresponding HOMFLY and Kauffman polynomials
are just their particular cases on particular hyperplanes of codimension one.
This fact opens absolutely new perspectives for study of both the universality
and the colored knot polynomials and deserves extension in various directions:
to other knots, to superpolynomials and to other representations from the
family of adjoint descendants.

Another application of present results can be in the study of the theory of Jacobi diagrams and Vogel's $\Lambda$-algebra. From the point of view of the Chern-Simons theory above polynomials are Wilson averages for a given knot in adjoint representation. In a given order in perturbation theory it is the sum of Jacoby diagrams (depending on gauge group) weighted with space-time integrals, independent of gauge group, but dependent on knot. So, for a finite set of Jacobi diagrams of given order we get 1-parameter universal values for their different combinations, where integer parameter runs values up to the order of diagram, approximately. Since the number of Jacobi diagrams is about the square of their order \cite{Vogel}, would we have 2-parameter universal expressions, we can hope to find universal expression to any particular Jacobi diagram, which should be compared with results of Vogel \cite{Vogel}.

\section{Acknowledgements}

AM's are grateful to Satoshi Nawata for a useful correspondence. RM is indebted to P.Deligne for discussions and e-mail exchange, which particularly includes \cite{D13,RMD}.

Our work is partly supported by grants NSh-1500.2014.2 (AM's) and 15-31-20832-Mol-a-ved (A.Mor.), by RFBR grants 13-02-00457 (A.Mir.) and 13-02-00478 (A.Mor.), by joint grants 15-52-50034-YaF (AM's), 15-51-52031-NSC-a (AM's), by 14-01-92691-Ind-a (AM's), by the Brazilian National Counsel of Scientific and Technological Development (A.Mor.), by Volkswagen Foundation (RM) and the Science Committee of the Ministry of Science and Education of the Republic of Armenia (RM).

\section{Appendix A}

Quantum dimensions for first few irreps of $SU(N)$ and $SO(N)$ are:
{\footnotesize
$$
\hspace{-1cm}\begin{array}{|c|c|}
\hline &\\
SU(N) \ \ {\rm with}\ A=q^N & SO(N)\ \ {\rm with} \ A=q^{N-1}\\
& \\
\hline &\\
 D_{[1]}=[N] = \frac{\{A\}}{\{q\}}& d_{[1]}=[N-1]+1 = \frac{\{A\}}{\{q\}}+1 =D_{[1]}(A,q)+1     \\ &\\
 \hline  & \\
D_{[2]} = \frac{[N][N+1]}{[2]} =
\frac{\{A\}\{Aq\}}{\{q\}\{q^2\}}& d_{[2]} = [N-1]\cdot\left(\frac{[N]}{[2]}+1\right)
=\left(1+\frac{\{q^2\}}{\{Aq\}}\right)\cdot D_{[2]}(A,q)
  \\
D_{[11]} = \frac{[N][N-1]}{[2]} = \frac{\{A\}\{A/q\}}{\{q\}\{q^2\}}&
d_{[11]}=[N-1]\cdot\left(\frac{[N-2]}{[2]}+1\right)
= \left(1+\frac{\{q^2\}}{\{A/q\}}\right)\cdot D_{[11]}(A,q)  \\ & \\
\hline  &\\
D_{[3]} = \frac{[N][N+1][N+2]}{[2][3]}= \frac{\{A\}\{Aq\}\{Aq^2\}}{\{q\}\{q^2\}\{q^3\}}
&d_{[3]}={[N][N-1]\over [2]}\left(1+{[N+1]\over [3]}\right)=\left(1+\frac{\{q^3\}}{\{Aq^2\}}\right)\cdot D_{[3]}(A,q)  \\
D_{[21]} = \frac{[N+1][N][N-1]}{[3]}= \frac{\{Aq\}\{A\}\{A/q\}}{\{q\}^2\{q^3\}}
&d_{[21]}=[N][N-2]\left(1+{[N-1]\over [3]}\right)=\left(1+\frac{\{q^3\}}{\{A\}}\right)\cdot D_{[21]}(A,q)  \\
D_{[111]} = \frac{[N][N-1][N-2]}{[2][3]}= \frac{\{A\}\{A/q\}\{A/q^2\}}{\{q\}\{q^2\}\{q^3\}}
&d_{[111]}={[N-1][N-2]\over [2]}\left(1+{[N-3]\over [3]}\right)=\left(1+\frac{\{q^3\}}{\{A/q^2\}}\right)\cdot D_{[111]}(A,q)  \\ & \\
\hline   &\\
D_{[4]} = \frac{[N][N+1][N+2][N+3]}{[2][3][4]} = \frac{\{A\}\{Aq\}\{Aq^2\}\{Aq^3\}}{\{q\}\{q^2\}\{q^3\}\{q^4\}}
&d_{[4]}={[N-1][N][N+1]\over [2][3]}\left(1+{[N+2]\over [4]}\right)=\left(1+\frac{\{q^4\}}{\{Aq^3\}}\right)\cdot D_{[4]}(A,q)  \\
D_{[31]} =  \frac{[N+2][N+1][N][N-1]}{[2][4]} =\frac{\{Aq^2\}\{Aq\}\{A\}\{A/q\}}{\{q\}^2\{q^2\}\{q^4\}}
&d_{[31]}={[N-2][N-1][N+1]\over [2]}\left(1+{[N]\over [4]}\right)=\left(1+\frac{\{q^4\}}{\{Aq\}}\right)\cdot D_{[31]}(A,q)  \\
D_{[22]} = \frac{[N+1][N]^2[N-1]}{[2]^2[3]}  = \frac{\{A\}^2\{Aq\}\{A/q\}}{\{q\}\{q^2\}^2\{q^3\}}
&d_{[22]}= {[N+1][N-3]\over [2]^2[3]}\Big(1+[N-1]\Big)\Big([3]+[N-1]\Big)= \\
&=\frac{\{Aq^2\}\{A/q^2\}}{\{q\}\{q^2\}^2\{q^3\}}\cdot(A+q)(1-{1\over Aq})(A+q^3)(1-{1\over Aq^3})   \\
D_{[211]} = \frac{[N+1][N][N-1][N-2]}{[2][4]} = \frac{\{Aq\}\{A\}\{A/q\}\{A/q^2\}}{\{q\}^2\{q^2\}\{q^4\}}
&d_{[211]}={[N][N-1][N-3]\over [2]}\left(1+{[N-2]\over [4]}\right)=\left(1+\frac{\{q^4\}}{\{A/q\}}\right)\cdot D_{[211]}(A,q)  \\
D_{[1111]}= \frac{[N][N-1][N-2][N-3]}{[2][3][4]}  = \frac{\{A\}\{A/q\}\{A/q^2\}}{\{q\}\{q^2\}\{q^3\}\{q^4\}}
&d_{[1111]}={[N-1][N-2][N-3]\over [2][3]}\left(1+{[N-4]\over [4]}\right)=\left(1+\frac{\{q^4\}}{\{A/q^3\}}\right)\cdot D_{[1111]}(A,q)  \\   & \\
\hline
\end{array}
$$
}

These dimensions satisfy the necessary sum rules:
\be
d_{[1]}^2 = d_{[2]}+d_{[11]} +1 \\
d_{[1]}^3 = d_{[3]}+2d_{[21]} + d_{[111]} + 3d_{[1]}   \\
d_{[1]}^4 = d_{[4]}+3d_{[31]}+2d_{[22]}+3d_{[211]} + d_{[1111]} + 6d_{[2]}+6d_{[11]}+3   \\
\ldots
  \\
d_{[2]}d_{[1]} = d_{[3]}+ d_{[21]} + d_{[1]}  \\
d_{[11]}d_{[1]} = d_{[21]} + d_{[111]} + d_{[1]}   \\
\ldots
  \\
d_{[2]}^2= d_{[4]}+d_{[31]}+d_{[22]} +d_{[2]}+d_{[11]}+1   \\
d_{[2]}d_{[11]} = d_{[31]}+d_{[211]}+ +d_{[2]}+d_{[11]}  \\
d_{[11]}^2 = d_{[22]}+d_{[211]}+d_{[1111]}+d_{[2]}+d_{[11]}+1   \\
\ldots
 \\
d_{[3]}d_{[1]} = d_{[4]}+d_{[31]} +d_{[2]}   \\
d_{[21]}d_{[1]} = d_{[31]}+d_{[22]} + d_{[211]} + d_{[2]}+d_{[11]}   \\
d_{[111]}d_{[1]}= d_{[211]}+ d_{[1111]}+d_{[11]}   \\
\ldots
\ee

\section{Appendix B}

As follows from (\ref{qdSO}), the quantum dimensions for $SO(N)$ algebras do not factorize in the variables $A,q$, but
they do so in $a=\sqrt{A}$, $Q=\sqrt{q}$, for example
\be
D_{Adj}^{SO(N)}=d_{[11]} =
\left(1+\frac{\{q^2\}}{\{A/q\}}\right) \frac{\{A\}\{A/q\}}{\{q\}\{q^2\}}
=\frac{\{aQ\}\{a^2\}\{a^2/Q^6\}}{\{a/Q^3\}\{Q^4\}\{Q^2\}}
\ee
which, in terms of $N$, is
\be
D_{Adj}^{SO(N)}= [N-1]\cdot\left(\frac{[N-2]}{[2]}+1\right)=\frac{[N/2][N-1][N-4]}{[N/2-2][2]}
\ee
and "looks better" for the series $D_N=SO(2N)$:
\be
D_{Adj}^{SO(2N)} = \frac{[N][2N-1][2N-4]}{[N-2][2]}
\ee
For exceptional algebras the factorized formulas are even more involved:
\be
D_{Adj}^{G_2} = \frac{[8/3][7/3][5]}{[4/3][5/3]}
\ \ \stackrel{q\longrightarrow 1}{\longrightarrow} \ \ 14,   \\
D_{Adj}^{F_4} = \frac{[10][13/2][6]}{[3][5/2]}
\ \ \stackrel{q\longrightarrow 1}{\longrightarrow} \ \ 52,   \\
D_{Adj}^{E_6} = \frac{[13][9][8]}{[4][3]}
\ \ \stackrel{q\longrightarrow 1}{\longrightarrow} \ \  78,   \\
D_{Adj}^{E_7} = \frac{[19][14][12]}{[6][4]}
\ \ \stackrel{q\longrightarrow 1}{\longrightarrow} \ \  133,   \\
D_{Adj}^{E_8} = \frac{[31][24][20]}{[10][6]}
\ \ \stackrel{q\longrightarrow 1}{\longrightarrow} \ \  248
\ee

Another dimension, contributing to the antisymmetric square, see (\ref{dimssquare}) below, is
\be
D_{X_2}^{G_2} = \frac{[6][5][11/3][7/3]}{[3][2][5/3][1/3]}
\ \ \stackrel{q\longrightarrow 1}{\longrightarrow} \ \ 77\ =\ \frac{14\cdot 13}{2}-14,   \\
D_{X_2}^{F_4} = \frac{[11][10][7][15/2][13/2][4][7/2]}{[11/2][5][5/2][2]^2[3/2]}
\ \ \stackrel{q\longrightarrow 1}{\longrightarrow} \ \ 1274\  =\ \frac{52\cdot 51}{2}-52,   \\
D_{X_2}^{E_6} = \frac{[14][13][10][9]^2[5]^2}{[7][5][3]^2[2]^2}
\ \ \stackrel{q\longrightarrow 1}{\longrightarrow} \ \  2925\ =\ \frac{78\cdot 77}{2}-78,   \\
D_{X_2}^{E_7} = \frac{[20][19][15[14][13]}{[5][4][3][2]}
\ \ \stackrel{q\longrightarrow 1}{\longrightarrow} \ \  8645\ =\ \frac{133\cdot 132}{2}-133,   \\
D_{X_2}^{E_8} = \frac{[32][31][25][24][21][18][14]}{[16][12][9][6][5][2]}
\ \ \stackrel{q\longrightarrow 1}{\longrightarrow} \ \  30380\ =\ \frac{248\cdot 247}{2}-248
\ee
and the non-trivial doublets, contributing to the symmetric square are:
\be
D_{Y_2}^{G_2} = \frac{[10/3][11/3][4][7]}{[4/3][5/3][2]}
\ \ \stackrel{q\longrightarrow 1}{\longrightarrow} \ \ 77,   \\
D_{Y_2}^{F_4} = \frac{[12][9][15/2][7][13/2][6]}{[2][5/2][3][7/2][4]}
\ \ \stackrel{q\longrightarrow 1}{\longrightarrow} \ \ 1053,   \\
D_{Y_2}^{E_6} = \frac{[15][12][10][9]^2[8]}{[5][4]^2[3][2]}
\ \ \stackrel{q\longrightarrow 1}{\longrightarrow} \ \  2430,   \\
D_{Y_2}^{E_7} = \frac{[21][18][15][14][13][12]}{[7][6][5][4][2]}
\ \ \stackrel{q\longrightarrow 1}{\longrightarrow} \ \  7371,   \\
D_{Y_2}^{E_8} = \frac{[33][30[25][24][21][20]}{[11][10][7][6][2]}
\ \ \stackrel{q\longrightarrow 1}{\longrightarrow} \ \  27000
\ee
\be
D_{Y_2'}^{G_2} = \frac{[5][4]}{[5/3][4/3][1/3]}
\ \ \stackrel{q\longrightarrow 1}{\longrightarrow} \ \ 27 = \frac{14\cdot 15}{2}-77-1,   \\
D_{Y_2'}^{F_4} = \frac{[10][9][15/2][6][3/2]}{[5][3][5/2][1/2]}
\ \ \stackrel{q\longrightarrow 1}{\longrightarrow} \ \ 324 = \frac{52\cdot 53}{2}- 1053-1,   \\
D_{Y_2'}^{E_6} = \frac{[13][12][10][8][5]}{[6][4]^2}
\ \ \stackrel{q\longrightarrow 1}{\longrightarrow} \ \  650 = \frac{78\cdot 79}{2}- 2430-1,   \\
D_{Y_2'}^{E_7} = \frac{[19][18][15][12]}{[5][4][2]}
\ \ \stackrel{q\longrightarrow 1}{\longrightarrow} \ \  1539 = \frac{133\cdot 134}{2}-7371-1,   \\
D_{Y_2'}^{E_8} = \frac{[31][30][25][20][14]}{[10][7][6][4]}
\ \ \stackrel{q\longrightarrow 1}{\longrightarrow} \ \  3875 = \frac{248\cdot 249}{2}-27000 - 1
\ee

In these formulas $Y_2, Y_2^{'}, Y_2^{''}$ are $Y_2(\alpha), Y_2(\beta), Y_2(\gamma)$, respectively, in  general decomposition (\ref{s2ad}). However, for exceptional algebras  $Y_2''=Y_2(\gamma)=0$.

Knot polynomials for exceptional algebras are rather lengthy, to give an example, we list them for the trefoil in sec.5.1.5.

\end{document}